\documentclass{aa}
\usepackage{soul}
\usepackage{comment}
\usepackage{xcolor}
\usepackage[table]{xcolor}
\usepackage{cuted}
\usepackage{hyperref}
\usepackage[normalem]{ulem}

\usepackage{xcolor}
\usepackage[dvipsnames]{xcolor}
\definecolor{darkred}{RGB}{210,0,0}

\usepackage{graphicx}
\usepackage{float} 
\usepackage{subcaption}
\usepackage{txfonts}

\begin{document}

   \title{Wolf-Rayet stars as tracers of gamma-ray emission}

   \subtitle{Isolated stars and stellar clusters/associations}

 \author{A. Inventar
          \inst{1}\fnmsep\thanks{Corresponding author; inventar@apc.in2p3.fr}
          \and
          G. Peron\inst{2}
          \and
          S. Recchia\inst{2,3}
          \and
          S. Gabici\inst{1}
          }

   \institute{Universit\'e Paris Cit\'e, CNRS, Astroparticule et Cosmologie, F-75013 Paris, France
         \and
             INAF Osservatorio Astrofisico di Arcetri Largo Enrico Fermi, 5, 50125, Firenze, Italy
        \and IFJ-PAN, Institute of Nuclear Physics Polish Academy of Sciences, PL-31342 Krakow, Poland}

   \date{Received 16/10/2025; Accepted 07/05/2026}

  \abstract 
   {Recent gamma-ray observations of young star clusters revealed that stellar wind termination shocks accelerate particles.
   The energy reservoir for particle acceleration is provided by the wind mechanical power of cluster member stars.
   }
   {
   Our goal is to provide a list of promising targets to guide future gamma-ray studies of stellar clusters and associations powered by massive stars.
   As the wind power of a single Wolf-Rayet (WR) star 
   can be of the same order than the cumulative wind power of the most massive star clusters we extend our study also to isolated WR stars.
   This is particularly interesting as a large fraction of WR stars are indeed isolated.
    }
   {We first ranked a large sample of star clusters and associations according to the number of member WR stars over their distance squared, a quantity which is proportional to the expected gamma-ray signal. 
   We then searched for possible spatial correlations between these objects and known gamma-ray sources.
   We repeated the same procedure for a sample of individual WR stars for which wind mechanical power and distance are known.
   }
   {We found a hint ($\lesssim$ 3~$\sigma$ confidence) for a correlation between star clusters hosting WR stars and unidentified GeV gamma-ray sources and we list new spatial associations for 11 clusters. Moreover, we found spatial coincidences between 4 isolated WR stars (WR110, WR114, WR111, and WR14) and unidentified gamma-ray sources. While we do not find a correlation between isolated WR stars and gamma-ray sources, we stress that these 4 isolated stars are characterised by extremely large values of the wind power divided the distance squared, which is a necessary condition to have gamma-ray emission detectable by current instruments.
   Assuming that the gamma-ray emission is powered by the WR stellar wind, we interpret it as the result of interactions between particles accelerated at the wind termination shock and ambient matter or photon fields. 
   Further studies on the most promising targets (both clusters and isolated WR stars) are encouraged in order to validate or confute the associations.
   }
   {Based on the fact that the wind power of an individual WR star can rival that of a full stellar cluster, we present a ranking of stellar clusters and isolated WR stars that may represent potential gamma-ray emitters.
   }


   \maketitle

\section{Introduction}
\label{sec:intro}

It was recognized long ago that stellar wind termination shocks (WTSs) might act as particle accelerators and contribute to the observed flux of cosmic rays (CRs) \citep{caspau1980}.
However, until very recently, no direct observational proof of this hypothesis was available.
As the wind mechanical power increases steeply with stellar mass, OB and Wolf-Rayet (WR) stars were promptly identified as promising sites for acceleration \citep{cesmon1983,volfor1982}.

Remarkably, the acceleration of the chemically enriched WR wind material might also explain the overabundance of $^{22}$Ne in CRs (\citealt{caspau1982}).
However, in order to reproduce the enhancement, accelerated wind material has to be significantly diluted in standard interstellar matter.
This points to the following scenario for CR origin: while most CRs are accelerated out of the interstellar medium at supernova remnant shocks, a small contribution (roughly at the 5-10\% level, \citealt{tatisc2021}) from chemically enriched stellar wind material is needed to account for the $^{22}$Ne excess.
Such additional component is most likely wind material accelerated at stellar WTSs \citep{tatisc2021}. \textit{
Thus, stellar WTS likely provide a minor but necessary contribution to Galactic CRs.}

The same conclusion can be reached by comparing the total mechanical energy injected by stellar WTSs and supernova explosions in the Galaxy.
This was done by \citet{cesmon1983} who found that stellar winds provide a comparatively small power and therefore, provided that the acceleration efficiency is similar at WTS and supernova remnant shocks, the latter provide the dominant contribution to the observed CR intensity. 

Pioneering searches of gamma-ray emission from the direction of WR stars can be found in \citet{Kaul_Mitra1997, Romero1999, Munar_Adrover2011}.
\citet{Kaul_Mitra1997} searched for spatial associations within WR stars and unidentified gamma-ray sources from the second EGRET catalogue, while \citet{Romero1999} used the third one.
The latter found a marginally significant correlation with massive stars (WR and OB associations), and listed as most promising gamma-ray sources WR~140, WR~142, and a binary system with a WR member in Cyg OB2.
More recently, \citet{Munar_Adrover2011} searched for associations between unidentified sources from the first Fermi/LAT catalogue and young massive stellar objects.
They found two associations with WR stars in the crowded Galactic center region, and no significant correlation with OB associations. Finally, \citet{Wang2022} found gamma-ray emission in the direction of a star cluster candidate hosting two WR stars.

Recently, the interest in stellar WTSs as particle accelerators was revived after several star clusters were identified in gamma-rays (see \citealt{gabici2024} or \citealt{PeronICRC2025} and references therein), and even proposed to be the accelerators of the highest energy Galactic CRs \citep{aharon2019}.
The gamma-ray emission can be ascribed to either hadronic \citep[e.g.][]{aharon2019} or leptonic \citep[e.g.][]{harer2023} interactions of CRs with ambient matter or radiation.
As star clusters may host not only stellar WTSs but also supernova shocks \citep[e.g.][]{gupta2020,vieu2022}, it is difficult to unambiguously determine the origin of the CR particles responsible for the gamma-ray emission.
However, the recent association of high-energy gamma-ray sources with very young star clusters embedded in the parent molecular cloud demonstrated that stellar winds do accelerate particles at least up to $\lesssim$~TeV energies \citep{peron2024, peron2025}.
This is because, given their young age (<~1~Myr), such clusters certainly did not witness a supernova explosion.
This leaves WTSs as the most likely  particle accelerators.
Moreover, the dense environment where such clusters are found suggests that the emission is hadronic \citep{Mestre2021, peron2024,peron2024b,peron2025}.
On the other hand, when clusters are found in comparatively low density regions, leptonic emission is likely to play a major role \citep[e.g.][]{harer2023}.

Any attempt to constrain the CR acceleration efficiency at WTSs from gamma-ray observations requires the knowledge of the global wind mechanical power of member stars, which is the energy reservoir for CR acceleration.
In order to estimate that quantity, a method based on the integration of stellar properties over the cluster initial mass function was developed by \citet{celli2024}.
This method provides an estimate for the global wind mechanical power of star clusters, but is limited by the uncertainty in the maximum mass of stars belonging to a given cluster, which may lead to underestimate the total power (i.e. the estimates provide lower limits for the wind mechanical power).

A crucial point is that, as the mechanical power of stellar winds increases very rapidly with stellar mass \citep[e.g.][]{seo2018}, the wind power of a star cluster is mostly due to the winds of very few most massive stars. Therefore, quite accurate estimates of the cluster wind power can be obtained in a reliable way by just summing the wind powers of such most massive stars \citep[e.g.][]{vieu2024}. As Wolf-Rayet (WR) stars blow the most powerful winds, we focus on these objects. They represent a late stage in the evolution of massive stars, which lasts a comparatively short time ($\tau_{WR} \sim 10^5$~yr, versus the millions of years that stars spend in main sequence) but is characterized by a release of wind mechanical energy roughly comparable to that released during the entire main sequence phase \citep{crowth2007,seo2018}. For this reason, they are promising targets in the search of radiative signatures of particle acceleration at WTS. In Sec.~\ref{sec:power} we discuss quantitatively this issue, and show that the wind of a single WR star can, in some cases, be as powerful as the integrated wind power from an entire star cluster. 

We also note that only stars with initial mass exceeding $\sim 25~M_{\odot}$ can go through the WR phase \citep{crowth2007}, and that their lifetime is shorter than $\approx 7$~Myr \citep[e.g.][]{limchi2006}.
Therefore, star clusters hosting WRs are most likely young.
This implies that, even though supernova explosions might contribute to the mechanical energy output of these systems (the most massive stars have a lifetime of $\approx$~3~Myr), the contribution from stellar winds is expected to be still important (if not dominant, for the youngest objects).
Star clusters hosting WR stars are then promising targets to study particle acceleration at WTS.
This is less true for very loose star clusters, often referred to as {stellar associations}, which may host several distinct stellar populations characterized by different ages. 
Nevertheless, also in this case searching for WR stars hosted in such systems is useful, because their number is related to both the global stellar wind power and the supernova explosion rate of the system.

To sum up, the goal of this paper is to provide a list of potentially bright gamma-ray emitters.
To do so, in Sec.~\ref{sec:rankclusters} we develop a method to rank both star clusters and stellar associations according to their global wind mechanical power, based on counting the number $N_{WR}$ of WR stars hosted by each cluster.
This ranking is used, then, to select the objects which are most promising target for gamma-ray observations.
Then, as most (more than 60\%) of the WR stars known to date do not belong to a cluster but are in fact isolated \citep{rate2020}, in Sec.~\ref{sec:rankWR} we investigate the possibility that the acceleration of CRs at the WTSs of such isolated stars might result in a detectable gamma-ray emission.
As done for clusters, we also rank individual WR stars according to their wind power.

To select the most promising gamma-ray emitters one should keep in mind that the gamma-ray flux from a given system depends on both stellar and environmental parameters, as well as the source distance $d$.
Stellar properties include the wind mechanical power $L_w$, i.e., the reservoir for CR acceleration, such that the observed flux depends on the combination $L_w/d^2$.
Relevant environmental parameters are the density of ambient matter and of soft stellar photons, which constitute the target for the CR interactions leading to the production of gamma-rays.
Powerful stars impact dramatically on their environment, inflating cavities in the interstellar medium, surrounded by a dense shell of compressed gas. 
These structures can be very complex and irregular \citep[see e.g. the simulations by][]{harer2025}. 
It is therefore not obvious to determine the appropriate target density for CR interactions.

For this reason, in Sec.~\ref{sec:rankclusters} and \ref{sec:rankWR} we will rank stellar systems according to the value of $L_w/d^2$ (or $N_{WR}/d^2$ for star clusters, assuming that this quantity is proportional to $L_w/d^2$) because a large value of $L_w/d^2$ is a { necessary} condition for bright gamma-ray emission.
We will then search for spatial associations between the best ranked objects and known and unidentified gamma-ray sources.
The combination of large value of $L_w/d^2$ and presence of gamma-ray emission of unknown origin in the region will define a list of objects which are potential gamma-ray emitters and that therefore deserve further attention (e.g. deeper observations, systematic searches of counterparts of the gamma ray emission, etc).

If confirmed, such associations would increase the number of known stellar cluster detected in gamma rays or, would establish isolated WR stars as a new class of gamma-ray sources.

\section{Mechanical power of stellar winds: individual WR stars versus entire star clusters}
\label{sec:power}

\begin{figure*}
   \centering
   \includegraphics[width=0.7\linewidth]{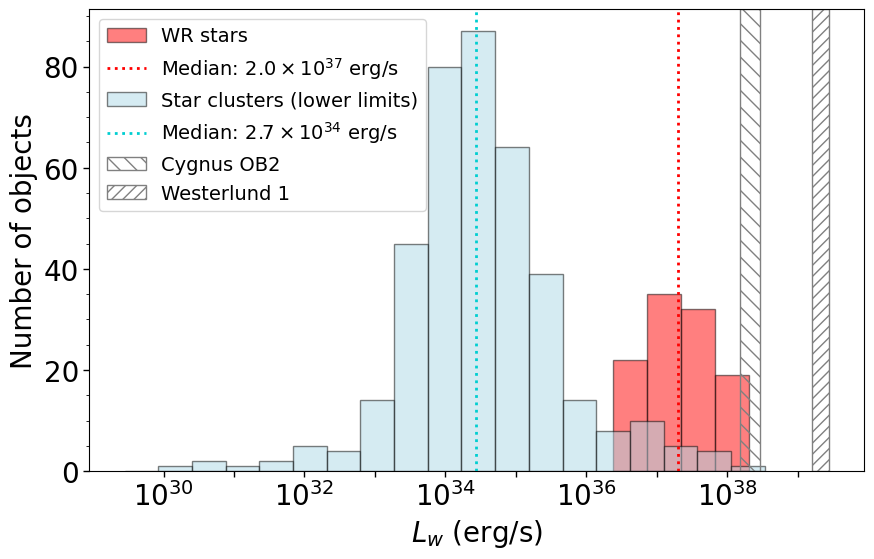}
   \caption{Distribution of wind powers for single WR stars (red) and entire star clusters (lower limits, from \citealt{celli2024}, cyan). Estimated wind powers for Cyg~OB2 \citep{menchi2024, vieu2024} and Westerlund~1 \citep{harer2023} are also shown.}
    \label{fig:histo}
 \end{figure*}

The energy reservoir for CR acceleration at the WTS of a WR star is the wind mechanical energy. 
Therefore, it is of paramount importance to obtain a reliable estimate of this quantity.
\citet{sander2019} used a model of stellar atmospheres to fit optical and UV stellar spectra and estimated the wind mass loss rate $\dot{M}$ and velocity $u_w$ of all putatively single Galactic WR stars of the carbon (WC) and oxygen (WO) sequence (or WR stars in binary systems for which the contribution to the spectrum from the companion star is negligible).
\citet{hamann2019} did the same for WR stars of the nitrogen (WN) sequence \citep[see][for a definition of spectral sequences]{crowth2007}.
In these works, the distance $d$ of WR stars was determined from parallaxes from the {Gaia} second Data Release (DR2).

The reliable measurement of $d$
makes it possible to estimate $\dot{M}$ and $u_w$ (typical value for WR stars are $\dot{M} \approx$ few times $10^{-5} M_{\odot}$/yr and $u_w \approx$ few times 1000~km/s), and therefore the wind power $L_w = (1/2) \dot{M} u_w^2$ with unprecedented accuracy.
This is done for 108 stars (the combined sample from both \citealt{sander2019} and \citealt{hamann2019}), i.e., 
about 15\% of all known Galactic WR stars \citep{roscro2015}\footnote{The compilation currently counts 709 WR stars, which is believed to be about half of all WR stars in the Milky Way, see 
\url{https://pacrowther.staff.shef.ac.uk/WRcat/index.php}}.

Before proceeding, it is worth reminding that other methods exist to estimate the mass loss rate and/or the wind terminal velocity of WR stars, for example based on radio observations \citep[e.g.][]{saha2023,blanco2024}.
These estimates are available from a relatively small number of WR stars, and may give in some cases estimates of the wind power that differ by a factor of several with respect to those by \citet{sander2019} and \citet{hamann2019} that we use here. 
In the following, when discussing individual objects, we will stress when different estimates of the wind power exist, and we will comment on possible implications.

The distribution of powers of WR stellar winds from \citet{sander2019} and \citet{hamann2019} is shown as a red histogram in Fig.~\ref{fig:histo}.
Powers span from a few times $10^{36}$ to $\gtrsim 10^{38}$~erg/s, with a median value of $\sim 2 \times 10^{37}$~erg/s. 
We compare that with the distribution of lower limits of the global wind powers estimated by \citet{celli2024} for star clusters identified using {Gaia} DR2.
They normalized the stellar mass distribution to the total number of stars observed by Gaia for each cluster and used empirical scalings to connect the stellar mass to $\dot{M}$ and $u_w$ to get the global wind power.
As they explain, in their approach the mass of the cluster is underestimated, and so is $L_w$.
In order to estimate how far these lower limits for $L_w$ are from the actual cluster wind powers, we show in the figure the estimate of the global wind power of Westerlund~1 (Wd1, $\sim 1.6-2.8 \times 10^{39}$~erg/s) as computed by \citealt{harer2023}. 
With a lower limit of $L_w > 3.15 \times 10^{38}$~erg/s, Wd1 is the most powerful among the star clusters studied by \citet{celli2024}. It is indeed the most powerful cluster in the Galaxy, rivaled only by the Nuclear star cluster in the Galactic center, and gamma-ray emission has been detected in its surroundings, almost coincident with the position of its collective wind termination shock \citep{aharon2022}.
The lower limit derived by \citet{celli2024} is a factor of $\Delta \sim 5-9$ smaller than the estimate by \citet{harer2023}. 
The global wind power is so large because the cluster hosts many massive stars (at least 166 stars of mass >~25~$M_{\odot}$ including 23 WR stars, \citealt{clark2020}).
We note that, even if we shift upward in $L_w$ the cyan histogram in Fig.~\ref{fig:histo} by a factor of $\Delta$, the power of single WR winds is still comparable to that of entire star clusters in the high-power end of the distribution. 

Apart from Wd1, another well studied system whose cocoon has been detected in gamma-rays is the Cygnus OB2 association \citep{ackerm2011,aharon2019, abeyse2021, lhaaso2024}. 
Its global wind power is $\sim 1.6 - 2.9 \times 10^{38}$~erg/s, and is dominated by three member WR stars and few massive O stars  \citep{menchi2024, vieu2024}.
This is just a factor of $\sim$~2 larger than the largest wind power of WR stars in our sample. 
We note that no estimate for the wind power of Cygnus OB2 was given by \citet{celli2024} as, being a quite loose stellar association, it does not appear in the list of stellar clusters identified in Gaia data. 

To conclude, the mechanical power of a single WR stellar wind can be comparable to the global wind power of the most massive clusters and, as we will see in the following, this fact can be used to select promising target (either star clusters/associations or isolated WR stars) for gamma-ray observations.

\section{Spatial association of star clusters hosting WR stars  and gamma-ray sources} 
\label{sec:rankclusters}

If a star cluster contains one or more WR stars, its global wind mechanical power will most  likely be dominated by those stars.
Here we assume that the number of
WR stars ($N_{WR}$) hosted in a cluster reflects his richness in terms of number of massive stars and is therefore roughly proportional to its global wind power, $L_w \propto N_{WR}$.
As seen in the previous Section, even a single WR star would provide a very large amount of mechanical energy, and therefore even clusters for which $N_{WR} = 1$ may be interesting targets for gamma-ray observations.

We further assume that a fraction $\eta$ of the wind power is converted into accelerated particles, so that the CR luminosity is $L_{CR} = \eta L_w$.
CRs accumulate in the system for a characteristic residence time $\tau_{res}$, which is the smallest between the time since the beginning of the CR injection, the CR confinement time in the region, and the CR energy loss time.
If the cluster is located at a distance $d$, the expected gamma ray emission is proportional to $F_{\gamma} \propto \eta L_{w} \tau_{res} \varrho_t/d^2$, where $\varrho_t$ represent the target density for CR interactions and is equal to the ambient gas number density for hadronic (neutral pion decay) gamma rays or to the energy density of the ambient soft radiation for inverse Compton gamma rays. 

As seen in the previous Section, estimates of the wind power of WR stars can be obtained with good accuracy for a number of stars, especially thanks to the estimate of their distances based on {Gaia} data.  
Unfortunately, the other physical quantities involved in the problem are often difficult to be constrained observationally.
For this reason, we introduce the quantity $\kappa \equiv \eta ~\tau_{res}~ \varrho_t$, which absorbs all these large uncertainties, to get $F_{\gamma} \propto \kappa (L_w/d^2)$.
This shows that a large value of $L_w/d^2$ is a necessary condition for a detection of a cluster in gamma rays, and that gamma-ray observations can be used to constrain the very uncertain value of the parameter $\kappa$ for those clusters for which $L_w$ and $d$ are reasonably well known.

In Table~\ref{tab:rankclusters} we list all the stellar clusters and associations that are known to host WR stars, according to Crowther's WR star catalog.
Objects are ranked according to the value $N_{WR}/d^2$, where the distance $d$ is taken equal to the average distance of the member WR stars.
In few cases, indicated with an asterisk, the distance of the cluster or association is taken from the literature (references are provided in the table caption). 
This was done whenever the dispersion of estimated WR star distances around the mean value is large. 
Also, in some cases we grouped clusters, as in the case of the three clusters Nuclear, Quintuplet, and Arches, which are located very close to each other, within few tens of parsecs from the Galactic center.
Clusters and/or associations have been grouped also when they host the same WR star.
This happens when a WR star belongs to a cluster which is in turn part of a looser stellar association.
In these cases, the number of WR stars associated to each component of the group/association are indicated in parentheses. 

As said above, \citet{celli2024} estimated the wind mechanical power for a large number of star clusters identified using {Gaia} data.
A comparison with their results is in order.
The main difference is that they started with a list of 390 clusters and computed their wind power, while our sample is composed by 44 clusters and associations only. These 44 stellar systems are known to host WR stars and are therefore good candidate gamma-ray emitter.
The overlap between the two samples consists of 12 objects only.
Naturally, these 12 objects are found within the 17 most powerful clusters in \citet{celli2024}.
Moreover, the six most powerful systems in their list all host WR stars.
They are Westerlund~1 (24 hosted WR stars), Danks~1 (6), Danks~2 (1), NGC~3603 (5), NGC~6231 (2) and Berkeley 87 (1).
On the other hand, our list includes several clusters which are not present in \citet{celli2024} (most notably the star clusters in the Galactic center region), and our method also deals with stellar associations (most notably the gamma-ray bright Cygnus OB2, and Carina OB1, which has been also proposed as a potential gamma-ray emitter, see references in Table~\ref{tab:rankclusters}).
Finally, Berkeley~86 and Dolidze~3 are amongst the 17 most powerful clusters in in their sample, but absent in ours. 
Both of them might host a WR star, but according to Crowther's catalog these associations are unlikely/questionable, and therefore we did not consider them.
We note however that Berkeley~86 belongs to the Cyg~OB1 association, which is in turn present in our list because it hosts other WRs.

The fact that a large fraction of the clusters in our list are not present in the one by \citet{celli2024} can be explained by at least three reasons.
First of all, loose stellar associations are not expected to be present in the {Gaia} catalogue.
Second, the clusters in our list from the VVV \citep{borissova2011} and GLIMPSE \citep{mercer2005} surveys have been discovered in the infrared, and might be too reddened to be seen by {Gaia}.
Third, some clusters (e.g. Dolidze~33 and Bochum~14) are present in the cluster catalogue of \citet{Hunt2024} based on {Gaia} DR3 data, but are not in earlier catalogues based on {Gaia} DR2, such as the one used by \citet{celli2024}.

We conclude then that the approach proposed here is complementary to that proposed by \citet{celli2024}.
It has the advantage of being very straightforward and to consider also loose stellar associations but the disadvantage of missing very young clusters (the WR phase begins towards the end of the star's lifetime, so that clusters younger than $\approx$~3~Myr are not expected to host WR stars); their relation to gamma-ray sources has been investigated by \citet{peron2024b}. 

In the following, for brevity, we will refer to the ensemble of stellar clusters and associations as {clusters}.

\subsection{Existing associations with gamma-ray sources}

\begin{table*}
\begin{center}
\caption{\centering Star clusters and stellar associations ranked according to $N_{WR}/\langle d \rangle ^2$, where $N_{WR}$ is the number of WR stars hosted by the cluster/association, and $\langle d \rangle$ the mean distance of these stars.}
    \begin{tabular}{| c | c | c | c | c | c |}
       \hline
    Cluster/{Association \tablefootmark{~a}} & $N_{WR}$ & $\langle {d_{kpc}} \rangle$ & $\frac{N_{WR}}{\langle d_{kpc} \rangle ^2}$ & $\gamma$-rays references & notes on $\gamma$-rays  \\
    \hline
    {Gamma Vel~\tablefootmark{~a}} & 1 & 0.336 & 8.9 & 1 & colliding wind binary \\
    \hline
    Westerlund~1 & 24 & 4.23\tablefootmark{*} & 1.3 & 2--5 & Likely \\
    \hline
    Galactic center& 76 & 8.00 & 1.2 & 4 & Tentative \\
    (Nuclear, Quintuplet, Arches) & (40,21,15) & & & & \\
    \hline
    {Sco OB1~\tablefootmark{~a}} & 3 & 1.56 & 1.2 & & \\
    (NGC~6231) & (2) &&&& \\
    \hline
    {Cyg OB2~\tablefootmark{~a}} & 3 & 1.64  & 1.1 & 4,6--10 & Likely \\
    \hline
    {Car OB1~\tablefootmark{~a}} & 5 & 2.50 & 0.8 & 11 & Tentative \\
    (Bochum~10, Collinder~228, Trumpler~16) & (1,1,1) & & & & \\
    \hline
    {Cyg OB1~\tablefootmark{~a}} & 3 & 2.02 & 0.74 &  & \\
     \hline
    {Dragonfish~\tablefootmark{~a}} & 10 & 4.00\tablefootmark{*} & 0.6 & & \\
    (Mercer~30, VVV~Cl011) & (4,1) & & & & \\
    \hline
    Mercer~20 & 2 & 1.95 & 0.53 & 12 &  Tentative \\ 
    \hline
    Collinder~121 & 1 & 1.42 & 0.50 & & \\
    \hline
    Danks~1 & 6 & 3.64 & 0.45 & 13 & Tentative  \\
    \hline
    Trumpler~27 & 2 & 2.16 & 0.43 & & \\
    \hline
    Berkeley~87 & 1 & 1.63 & 0.38 & 14,15 & Tentative \\
    \hline
    {Cen OB1~\tablefootmark{~a}} & 1 & 1.87 & 0.29 & & \\
    \hline
    {Cep OB1~\tablefootmark{~a}} & 4 & 3.87 & 0.27 & & \\
    \hline
    Cl~1813-178 & 2 & 2.82 & 0.25 & 16 & Tentative \\
    \hline
    VVV~Cl036 & 1 & 2.00 & 0.25 & & \\
    \hline
    HM~1 & 2 & 3.08 & 0.21 & & \\
    ({Anon Sco OB~\tablefootmark{~a}}) & (1) & & & & \\
    \hline
    VVV~Cl099 & 3 & 4.00\tablefootmark{*} & 0.19 & & \\
    \hline
    {Cyg OB3~\tablefootmark{~a}} & 1 & 2.36 & 0.18 & & \\
    \hline 
    {Cas OB1~\tablefootmark{~a}} & 1 & 2.40 & 0.17 & & \\
    \hline
    Dolidze~33 & 1 & 2.72 & 0.13 & & \\
    \hline
    Bochum~14 & 1 & 2.74 & 0.13 & &  \\
    \hline
    Markarian~50 & 1 & 2.80 & 0.13 & & \\
    \hline
    VVV~Cl073 & 2 & 4.00 & 0.13 & & \\
    \hline 
    Mercer~23 & 1 & 2.93 & 0.12 & & \\
    \hline 
    VVV~Cl074 & 4 & 6.00 & 0.11 & & \\
    \hline
    Westerlund~2 & 2 & 4.38 & 0.10 & 17-21 & Likely \\
    \hline
    {Cir OB1~\tablefootmark{~a}} & 1 & 3.20 & 0.098 & & \\
    (Pismis~20) & (1) & & & & \\
    \hline
    NGC~3603 & 5 & 7.18 & 0.097 & 22-24 & Likely \\
    \hline 
    Hogg~15 & 1 & 3.55 & 0.097 & & \\
    \hline
    Quartet & 3 & 6.30\tablefootmark{*} & 0.076 & & \\
    \hline
    VVV~Cl041 & 1 & 3.86 & 0.067 & & \\
    \hline
    Mercer~81 & 8 & 11.0 & 0.066 & 25 & Tentative \\
    \hline
    Ruprecht~44 & 1 & 4.18 & 0.057 & & \\
    \hline
    SGR~1806-20 & 4 & 8.70 & 0.053 & 26,27 & Tentative \\
    \hline
    [DBS2003] 179 & 3 & 7.90 & 0.048 & & \\
    \hline
    C1104-610a & 2 & 6.51 & 0.047 & & \\
    \hline 
    Danks~2 & 1 & 5.16 & 0.038 & 13 & Tentative  \\
    \hline
    { Bochum~7} & 1 & 5.25 & 0.036 & & \\
    \hline
    W43 cluster & 1 & 7.00 & 0.020 & 19,28-30 & Likely \\
    \hline
    Sher~1 & 1 & 7.13 & 0.020 & & \\
    \hline
    VVV~Cl009 & 1 & 7.68 & 0.017 & & \\
    \hline
    Mercer~70 & 1 & 7.90 & 0.016 & & \\
    \hline
    \end{tabular}
    \tablefoot{
    \tablefoottext{a}{Stellar associations}
    \tablefoottext{*}{Distances from: \citet{neguer2022}, \citet{sanche2024}, \citet{chene2013}, \citet{messineo2009}.}}
    \tablebib{ (1) \citet{martid2020}; (2) \citet{abramo2012}; (3) \citet{ohm2013}; (4) \citet{aharon2019}; (5) \citet{aharon2022}; (6) \citet{ackerm2011}; (7) \citet{abeyse2021}; (8) \citet{cao2021}; (9) \citet{astias2023}; (10) \citet{lhaaso2024}; (11) \citet{ge2022}; (12) \citet{Sun2022}
    (13) \citet{liu2024}; (14) \citet{mancha1996}; (15) \citet{bednar2007}; (16) \citet{guo2024}; (17) \citet{aharon2007}; (18) \citet{Abramowski2011}; (19) \citet{lemoin2011}; (20) \citet{Yang2018}; (21) \citet{Mestre2021}; (22) \citet{Yang2017}; (23) \citet{Saha2020}; (24) \citet{Rocamora2025}; (25) \citet{davies2012}; (26) \citet{Yeung2016}; (27) \citet{Hess2018_sgr}; (28) \citet{hessco2018}; (29) \citet{Yang2020}; (30) \citet{Cao2025}.}
    \label{tab:rankclusters}
\end{center}
\end{table*}

The last two columns of Table~\ref{tab:rankclusters} indicate likely or tentative associations between clusters that host at least 1 WR star and gamma-ray sources that have been reported in the literature.
The associations fall into three categories, that we describe below.
\begin{enumerate}
    \item{{\textit{Source powered by a single WR star --}} The cluster Gamma Velorum is characterised by the largest value of $L_w/\langle d \rangle ^2$, mainly due to its proximity to us. 
    An association with the Fermi source 4FGL~J0809.5-4714 has been proposed \citep{Pshirkov2016,martid2020}. 
    In fact, the position of the Fermi source and the hints of periodic variability of the gamma-ray emission point towards an association not with the cluster itself, but rather with the binary system $\gamma^2$ Velorum, hosted by the cluster and formed by a WR and a O star \citep{martid2020}. 
    The WR stellar wind dominates the total mechanical energy output of the system \citep{desark2025}, and this supports one of the ideas pushed forward in this paper, i.e., that the wind of a single WR star suffices to power a gamma-ray source. 
    We will discuss this extensively in Sec.~\ref{sec:rankWR}.
    }
    
    \item{{\textit{Likely associations with clusters --}} With likely association between a cluster and a gamma-ray source we mean here associations with sources from catalogues for which other plausible origins of the emission (pulsars, supernova remnants not belonging to the cluster, etc.) have been searched for but not identified.
    Following this criterion, there are 4 likely associations: Westerlund~1 (associated with HESS~J1646-458, \citealt{aharon2022}), Westerlund~2 (associated with 4FGL~J1023.3-5747e, \citealt{abdoll2020}, and with HESS~J1023-575, \citealt{aharon2007,Mestre2021}), NGC~3603 (associated with 4FGL J1115.1-6118, \citealt{Saha2020}), and W43 (associated with LHAASO J1848-0153u, \citealt{Cao2025}). 
    Finally, we add to the list a fifth object: Cyg OB2.
    The Cygnus region is extremely crowded, and contains both discrete gamma-ray sources (TeV J2032+4130, LHAASO J2032+4102) and extended diffuse emission.
    It is therefore difficult to identify the source(s) of the high-energy particles emitting gamma-rays.
    However, the very large wind mechanical power of Cyg OB2 makes it a prime candidate to explain particle acceleration in the region and the associated extended gamma-ray emission \citep{menchi2024,vieu2024,astias2023}.
    It is also possible that part of the gamma-ray emission is due to particle accelerated at the remnant of a supernova that exploded within the Cyg OB2 association \citep{harer2025}, and some contamination might also be expected from the X-ray binary Cygnus X-3, which contains the WR star WR~145a \citep{vankerk1992}.}
    \item{{\textit{Tentative associations with clusters --}} Tentative associations have been previously proposed (see Table~\ref{tab:rankclusters} for references) for 9 more objects in our list. 
    In some cases, associations with discrete sources have been proposed (Berkeley~87 associated with MGRO~J2019+37, 3EG~J2021+4716 and 3EG~J2016+3657; Cl~1813-178 with HESS~J1813-178 and 4FGL~J1813.1-1737e; Mercer~81 with HESS~1640-465; SGR~1806-20 with HESS~J1808-204 and 3FGL~J1809.2-2016c), while in other cases an extended  diffuse emission has been observed around the cluster/association (Galactic center, Car~OB1, Mercer~20, Danks~1 and 2).
    In these cases, other possible sources of the gamma-ray emission have been identified (see Appendix \ref{sec:appendix} for a discussion of this issue).}
\end{enumerate}

The five objects listed above as likely associations with gamma rays have been extensively studied, and estimates of their global wind power have been reported (see e.g. \citealt{harer2023} for Westerlund~1, \citealt{menchi2024} for Cyg~OB2, \citealt{drisse1995} or \citealt{Saha2020} for NGC~3603, \citealt{Mestre2021} for Westerlund~2, and \citealt{Yang2020} for W43).
Wind powers increase with $N_{WR}$, validating our ranking according to values of $N_{WR}/d^2$.
Remarkably, these five gamma-ray-bright objects are characterized by values of $N_{WR}/\langle d \rangle ^2$ spread over almost two decades.
This suggests that a large value of $N_{WR}/\langle d \rangle ^2$ (and therefore of $L_w/d^2$) is  not the only condition leading to the production of gamma rays.
The absence of gamma-ray emission for systems characterized by a large value of $N_{WR}/\langle d \rangle ^2$ can be explained in two ways.
Either the WR stars hosted by the cluster are not powerful enough (powers of individual WR stars span over $\sim$~2 orders of magnitudes, see Fig.~\ref{fig:histo}), or the quantity $\kappa$ defined above assumes a small value.
Given the large uncertainties in the determination of ambient physical parameters, $\kappa$ certainly plays a major role here. 
$\kappa$ can be small because either $\varrho_t$ is very small and there is no target to produce gamma rays, or the environment is not favorable to particle acceleration and therefore $\eta$ is very small, or the topology of the ambient magnetic field is such that CR do not accumulate within or close to the system ($\tau_{res}$ is very short).

\subsection{Selection criteria for new associations}
\label{3.2}

The five likely associations between clusters and gamma-ray sources can be used to define a criterion to search for further associations. 
To this purpose, let us briefly summarize  the characteristics of their gamma-ray emission, that can also be retrieved in Table \ref{tab:likely}.
Note that in the following we take as coordinates of each cluster the mean of the coordinates of its member WR stars.

\begin{itemize}
    \item{Westerlund~1 -- The GeV emission reported by \citet{ohm2013} is $\lesssim$~1 degree ($\lesssim 70$~pc at a distance of 4.23~kpc) offset from the cluster and has an extension (radius) of about half degree. 
    This emission is probably even more extended ($\sigma = 0.71^{\circ}$ or $\sim $52~pc), its centroid more offset from the cluster ($\sim$1.5$^{\circ}$ or $\sim$110~pc), and possibly associated to an outflow emanating from Westerlund~1 and carrying away CRs \citep{Lemoine_2025}.
    In the TeV domain, Westerlund~1 is associated to the source HESS~J1646-458 from the H.E.S.S. Galactic Plane Survey (HGPS, \citealt{hessco2018}). The distance between the source centroid and the cluster is $\sim$~0.45$^{\circ}$ ($\sim$~33~pc), and the source extension is $\sigma \sim 0.5^{\circ}$ ($\sim$37~pc).}
    \item{Cyg OB2 -- This stellar association is embedded in a very extended (several degrees) region, called Cygnus Cocoon, characterised by a diffuse gamma-ray emission in both GeV and TeV domain \citep{abeyse2021,astias2023}. 
    The extended source 4FGL J2028.6+4110e from the fourth Fermi/LAT catalogue \citep[4FGL, ][]{ballet2023} is associated to such emission, and its centroid is located $\sim 0.14^{\circ}$ ($\sim$~4.2 pc, at a distance of 1.64~kpc) away from the cluster.
    A very extended (>100 squared degrees) multi-TeV emission has also been revealed by LHAASO in this very crowded region \citep{lhaaso2024}. The source from the First LHAASO Catalog of Gamma-Ray Sources \citep{cao2024} which is the closest to Cyg~OB2 (and claimed to be possibly associated to this object, \citealt{cao2021}) is LHAASO~J2032+4102, at a distance of $\sim$~0.7$^{\circ}$ and 0.5$^{\circ}$ ($\sim$~19 and 14~pc, for LHAASO KM2A and WCDA array, respectively).}
    \item{Westerlund~2 -- It is associated to the extended Fermi source 4FGL~J1023.3-5747e and is coincident with the HGPS source HESS~J1023-575. The distance from the centroid of the emission of these sources is 0.11$^{\circ}$ ($\sim$~8~pc, for a distance of 4.83~pc) and 0.12$^{\circ}$ ($\sim$~9~pc), respectively.
    Also in this case, a very extended (several degrees) diffuse emission surrounding the cluster has been reported, based on Fermi/LAT observations~\citep{Yang2018}.}
    \item{NGC~3603 -- This cluster located in a dense star forming region has recently been associated to the Fermi unidentified source 4FGL~J1115.1-6118 \citep{Saha2020}, whose centroid is only 0.05$^{\circ}$ ($\sim$~6~pc at 7.2~kpc) from its center. 
    Extended GeV diffuse emission (of radius 1.1$^{\circ}$) has also been studied in this region \citep{Yang2017} but a source extension was later disputed by \citealt{Saha2020} and \citealt{Rocamora2025}. 
    No gamma-ray counterpart in the TeV domain has been reported.}
    \item{W43 cluster -- The eponymous cluster in this very active region is not associated with any 4FGL source. Nevertheless, diffuse emission (0.6$^{\circ}$) has been observed around W43 by \citet{Yang2020} and is thought to be linked to the cluster. 
    Moreover, in the TeV/multi-TeV domain, W43 has been associated to both H.E.S.S. and LHAASO sources. 
    The distance from the centroid of HESS~J1848-018 \citep{hessco2018} is 0.23$^{\circ}$ ($\sim$ 28 pc at 7kpc) while for LHAASO~J1848-0153u \citep{Cao2025} it is 0.21$^{\circ}$ ($\sim$ 26 pc) for KM2A and 0.17 $^{\circ}$ ($\sim$ 21 pc) for WCDA.
    A possible association with the HAWC source 2HWC~J1847-018 has also been reported \citep{abeyse2017}.}
\end{itemize}

\begin{table}[]
    \centering
    \caption{\centering Sources from the 4FGL, HGPS, and First LHAASO Catalogue (KM2A) which have been associated with the stellar clusters listed in the first column.}
    \begin{tabular}{l|c | c |  c | c |}
       Object  & $\vartheta$/$^{\circ}$ ($R$/pc) & Ext.    & Source \\
       \hline
        Westerlund~1 & 0.45 (33) & X & HESS~J1646-458 \\
        Cyg OB2 & 0.14 (4.2) & X & 4FGL~J2028.6+4110e \\
        & 0.66 (19) & - & 1LHAASO~J2031+4052 \\
        Westerlund~2 & 0.11 (8.1) & X & 4FGL~J1023.3-5747e \\
        & 0.12 (9.2) & X & HESS~J1023-575 \\
        NGC~3603 & 0.05 (5.8) & - & 4FGL~J1115.1-6118 \\
        W43 cluster & 0.23 (28) & X & HESS~J1848-018 \\
        & 0.21 (26) & X & 1LHAASO~J1848-0153u\\
    \end{tabular}
    \tablefoot{The second column indicates the distance between the centroid of the gamma ray emission and the position of the cluster in degrees (and parsecs). Extended (point like) sources are identified by an X (-) in the third column. The source catalog name is reported in the fourth column.}
    \label{tab:likely}
\end{table}

The diffuse emission detected around most of the clusters discussed above is interpreted as the result of CR escape from such systems, and of their interactions with ambient matter on quite large spatial scales.
In general, the detection of such diffuse emission extending up to several degrees requires a dedicated analysis of gamma-ray data.
On the other hand, if we limit ourselves to considering discrete gamma-ray sources, we see that in the cases considered above the projected distance $R$ between the cluster and the centroid of the gamma-ray emission varies between a few parsecs and $\sim $~30~pc (see Table~\ref{tab:likely}, where all catalogued sources associated to the clusters considered here are listed). 
This offset between gamma-ray emission and position of the stars is not surprising.
It is connected to the fact that stellar clusters and associations inject a substantial amount of mechanical energy which is then dissipated at both stellar WTS and at supernova remnant shock.
These shocks are characterized by sizes which are much larger than the extension of stellar systems \citep[e.g.][]{weaver1977, gupta2020,vieu2022} and therefore, depending on the spatial distribution of target material in the region, the gamma-ray emission can be offset/asymmetric with respect to the location of the cluster.
More quantitatively, on can use the theory of interstellar bubbles \citep[e.g.][]{weaver1977,gabici2023} to estimate the radius of the WTS as:
\begin{equation}
R_{WTS} \sim 30 ~ L_{38}^{3/10} n_1^{-3/10} u_3^{-1/2} t_1^{2/5}~\rm pc
\end{equation}
where $L_{38} = L_w/(10^{38} {\rm erg/s})$ is the total wind power of the system, $n_1 = n/({10~ \rm cm^{-3}})$ the ambient density, $u_3 = u_w/(10^3~{\rm km/s})$ the wind velocity, and $t_1 = t/(10~{\rm Myr})$ the system age.
The radius of the WTS of the most powerful isolated WR stars is slightly smaller, but still of the same order of magnitude \citep[][]{vanmarle2015}.

In the light of what is said above, we suggest that a good strategy to find new associations could be to search for gamma-ray sources within a projected distance of $d_{proj} \le 30$~pc from the cluster.
This corresponds to an angular distance equal to $\vartheta_{30} \sim 1.7^{\circ} (d/{\rm kpc})^{-1}$ around the clusters.
In fact, in some cases stars in clusters can be quite loose, i.e., spread over an extended region in the sky.
To estimate the extension of the region where most wind mechanical energy is injected, we compute for each object in Table~\ref{tab:rankclusters} the maximum angular distance $\vartheta_{max}$ of member WR stars from the cluster/association center. 
We perform then searches for gamma-ray sources within an angular region of radius $\vartheta_s = (\vartheta^2_{30} + \vartheta^2_{max})^{1/2}$.

\subsection{Search for spatial associations}
\label{sub:search}

\begin{figure*}[ht]
    \centering
    \begin{subfigure}[t]{0.33\textwidth}
        \centering
        \includegraphics[width=\linewidth]{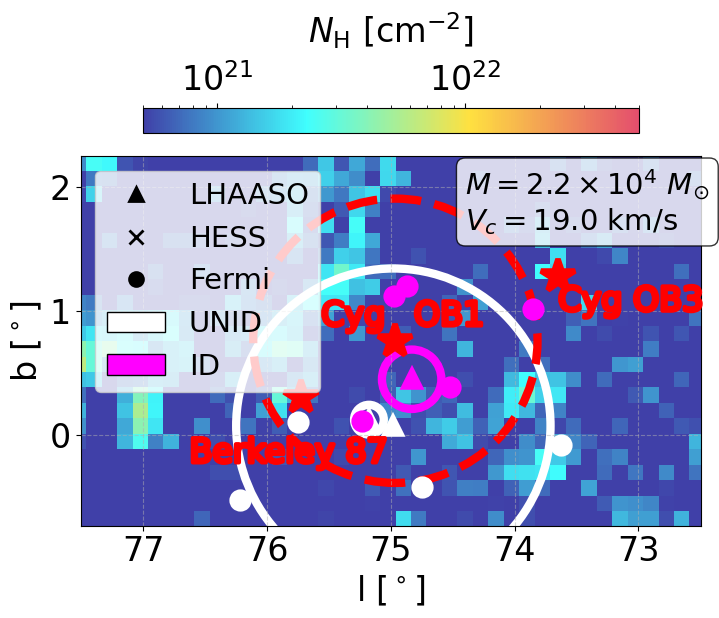}
    \end{subfigure}
    \hspace{1mm}
    \begin{subfigure}[t]{0.30\textwidth}
        \centering
        \includegraphics[width=\linewidth]{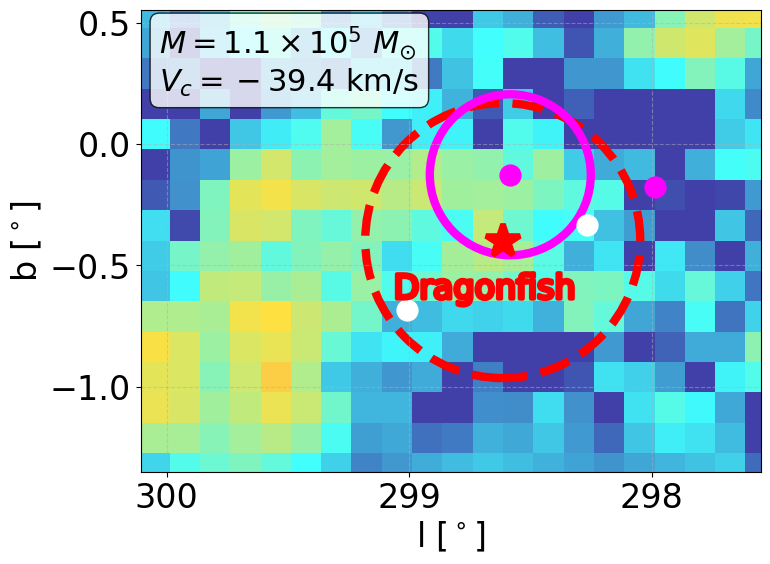}
    \end{subfigure}
    \begin{subfigure}[t]{0.30\textwidth}
        \centering
        \includegraphics[width=\linewidth]{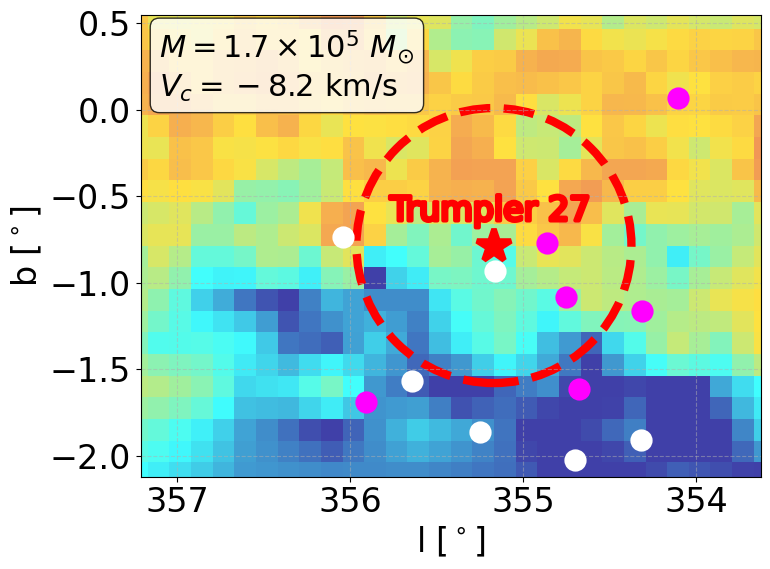}
    \end{subfigure}
    
    \centering
    \hspace{2mm}
    \begin{subfigure}[t]{0.3\textwidth}
        \centering
        \includegraphics[width=\linewidth]{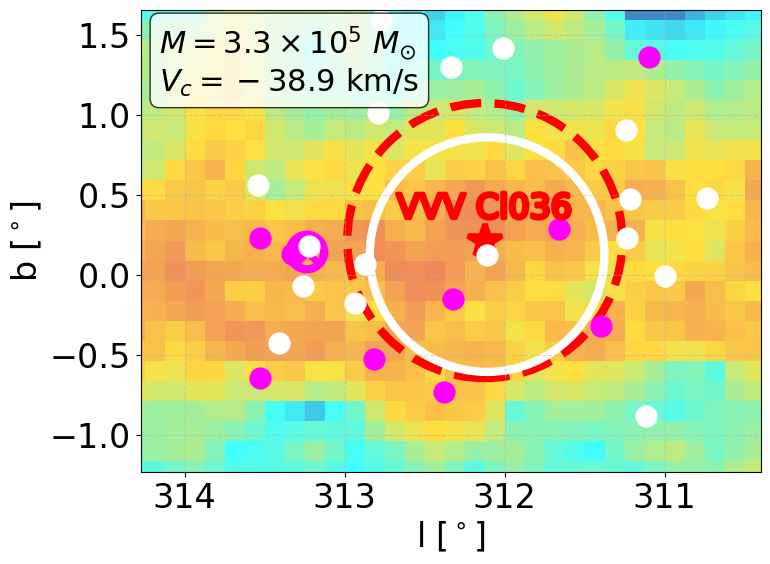}
    \end{subfigure}
    \hspace{3mm}
    \begin{subfigure}[t]{0.3\textwidth}
        \centering
        \includegraphics[width=\linewidth]{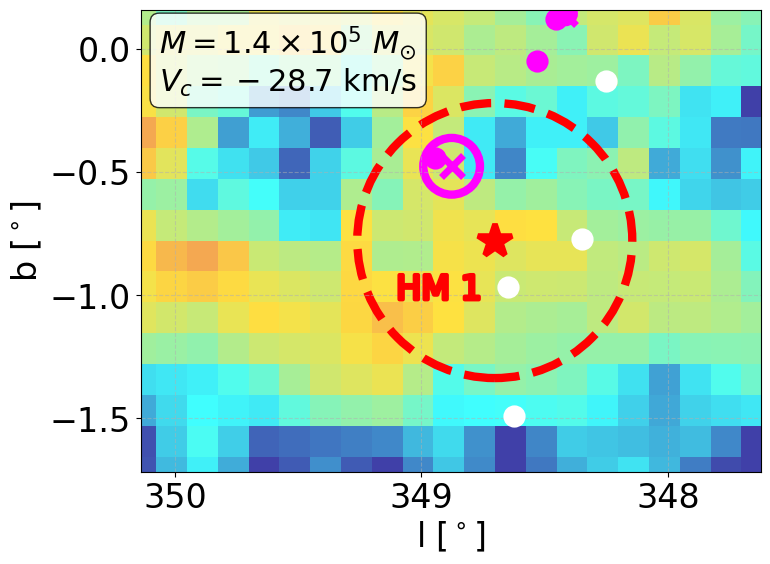}
    \end{subfigure}
    \begin{subfigure}[t]{0.3\textwidth}
        \centering
        \includegraphics[width=\linewidth]{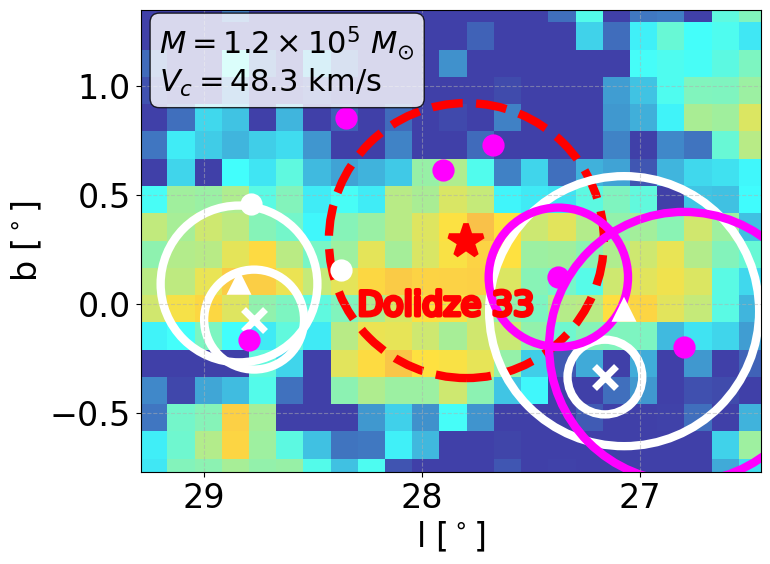}
    \end{subfigure}

    \centering
    \hspace{2mm}
    \begin{subfigure}[t]{0.3\textwidth}
        \centering
        \includegraphics[width=\linewidth]{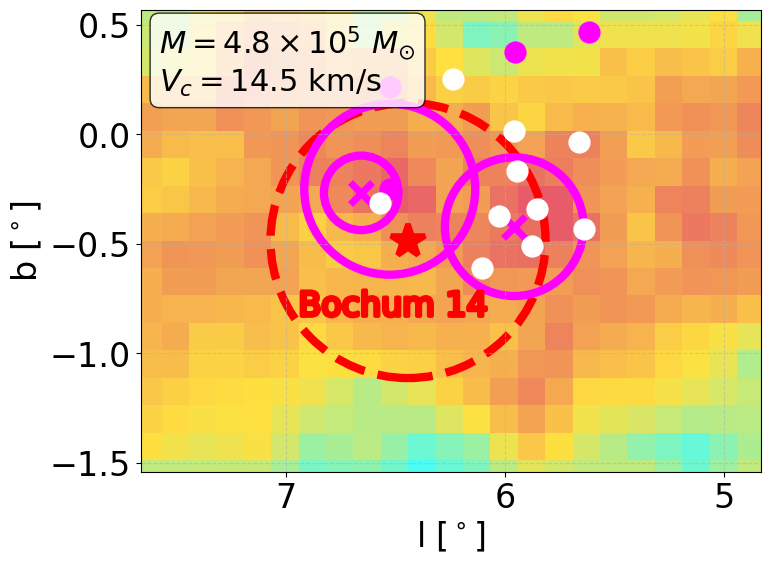}
    \end{subfigure}
    \hspace{3mm}
    \begin{subfigure}[t]{0.3\textwidth}
        \centering
        \includegraphics[width=\linewidth]{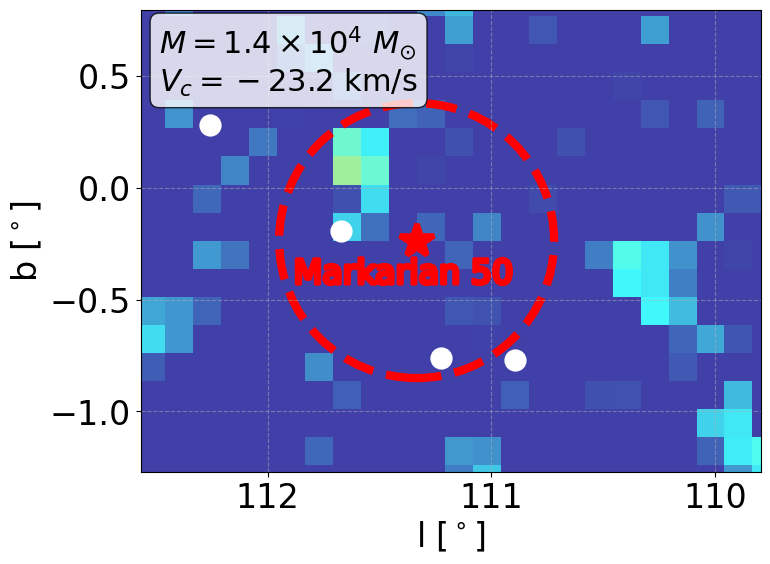}
    \end{subfigure}
    \begin{subfigure}[t]{0.3\textwidth}
        \centering
        \includegraphics[width=\linewidth]{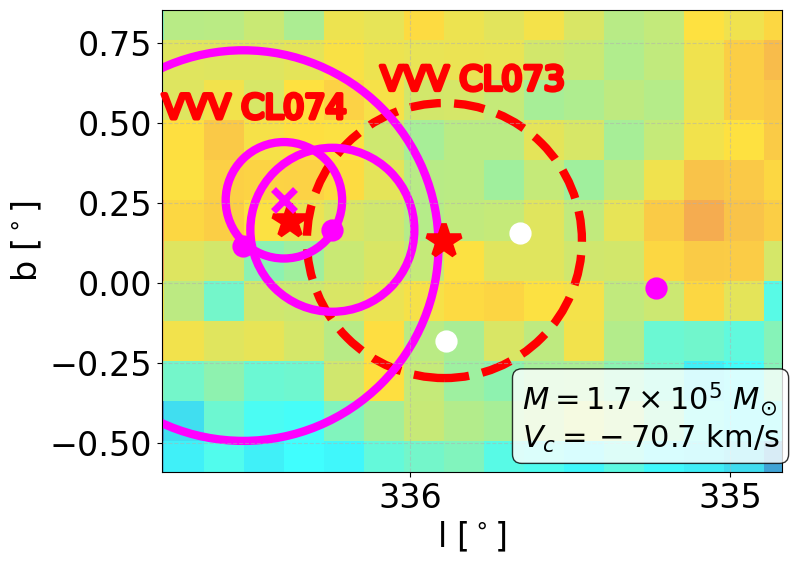}
    \end{subfigure}

    \hspace{-5.5cm}
    \begin{subfigure}[t]{0.305\textwidth}
        \centering
        \includegraphics[width=\linewidth]{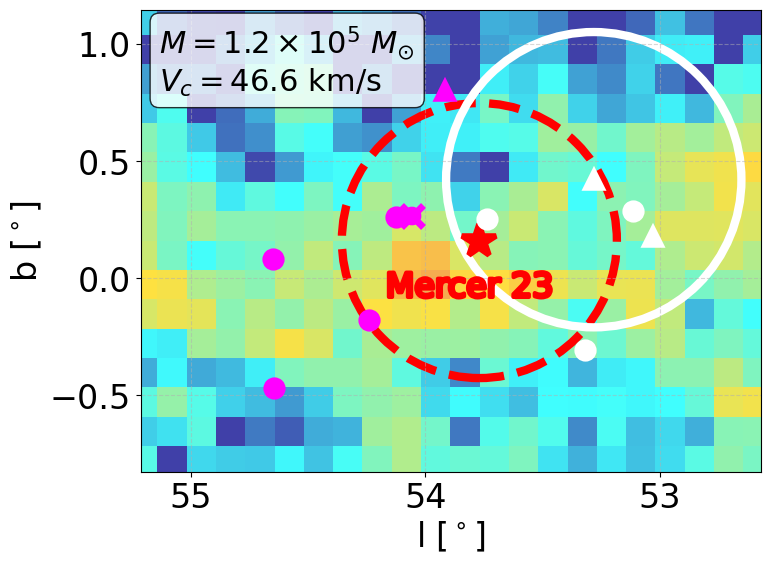}
    \end{subfigure}
    \hspace{3mm}
    \begin{subfigure}[t]{0.3\textwidth}
        \centering
        \includegraphics[width=\linewidth]{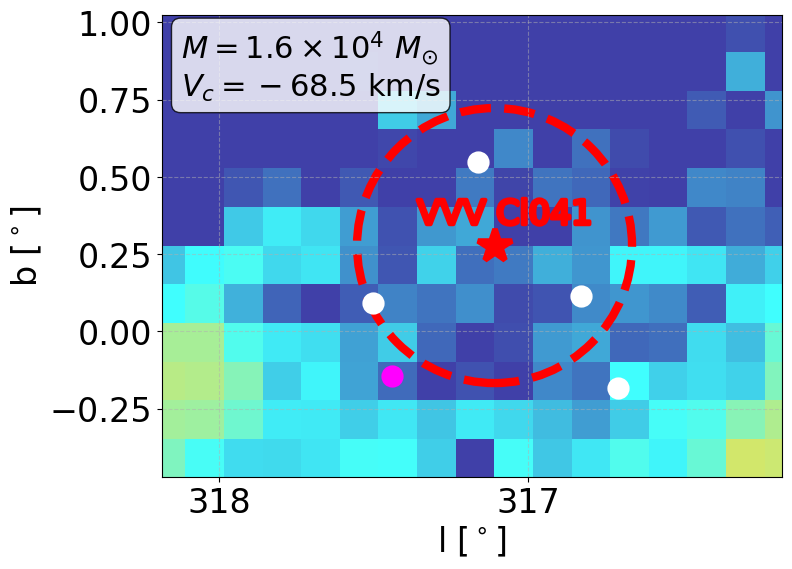}
    \end{subfigure}
    
    \caption{Position (magenta/white points) of gamma-ray sources found in the vicinity of the clusters from Table~\ref{tab:rankclusters} for which a spatial association with unidentified gamma-ray sources has been found. If the source is not point-like, its extension is indicated with a circle. The center of the cluster is indicated by a star and the red dashed circle shows the search region (see text for more details).The background color indicates the hydrogen gas column density, integrated in a range of $\pm 10 \rm ~ km/s$ around the velocity $V_c$ corresponding to the distance of the cluster. The total hydrogen mass in the search region is also indicated.}
    \label{fig:maps}
\end{figure*}

We performed systematic searches for spatial associations within the fourth Fermi/LAT catalog \citep[4FGL DR4, ][]{ballet2023}, the H.E.S.S. Galactic Plane Survey (HGPS, \citealt{hessco2018}), and the First LHAASO Catalog of Gamma-Ray Sources \citep{cao2024}. 
For HGPS, we updated the sources types with the most recent status (based on TeVCat, \citealt{wakhor2008}), to ensure that sources currently identified are not still treated as unidentified. For LHAASO, we consider the KM2A extension whenever the source is detected by KM2A, and the WCDA extension otherwise.
We considered the 29 objects in Table~\ref{tab:rankclusters} for which no association with gamma-ray sources (either likely or tentative) has been previously reported in the literature (i.e. entries in Table~\ref{tab:rankclusters} for which the 5$^{th}$ and 6$^{th}$ columns are empty). 
Among these objects, only 22 are within the part of the sky surveyed by the HGPS, and 8 within the field of view of LHAASO.
For 18 of the 29 objects considered, no unidentified gamma-ray sources were found within a distance $\vartheta_s$.
All of the remaining 11 objects are spatially associated with unidentified 4FGL sources and are either in the field of view of LHAASO or within the HGPS boundaries (or in both). 
Only two of these 11 objects are associated with LHAASO unidentified sources, while no spatial association have been found with unidentified HGPS sources.
The fact that for a significant fraction of the objects no gamma-ray emission has been found (despite the large values of $N_{WR}/\langle d \rangle ^2$) supports the idea that environmental effects likely play a crucial role in determining the gamma-ray output from such systems. 

Fig.~\ref{fig:maps} shows the maps of the regions surrounding the 11 clusters for which associations have been found.
Clusters of our sample (Table~\ref{tab:rankclusters}) are indicated with a red star, and the search region of size $\vartheta_s$ is indicated with a dashed red circle.
Fermi/LAT (4FGL), H.E.S.S. (HGPS), and First LHAASO catalog sources are indicated  with dot, cross and triangle symbols respectively.
Whenever the source is extended, its size is indicated with a circle (39\% containment region for H.E.S.S. and LHAASO sources, source extension for Fermi/LAT sources as reported in the 4FGL).
White (pink) colors refer to unidentified (identified) gamma-ray sources.
A table listing all sources found within the search region can be found in Appendix~\ref{tab:associationsALL}.

The background color in Fig.~\ref{fig:maps} indicates the ambient gas (hydrogen) column density, as inferred from $^{12}$CO line intensity \citep{Dame2001}, using a CO to H$_2$ conversion factor of 2$\times 10^{20}$~K$^{-1}$~km$^{-1}$~s \citep{bolatto2013}. 
The hydrogen column density ($N_{\rm H} = 2 \times N_{\rm H_2}$) has been derived integrating the CO emission over a velocity interval of $\pm 10$~km/s around the central velocity of the target position (derived using the Galactic rotation curve of \citealt{Mroz2018}), since this scale is comparable to the typical turbulent velocity dispersion of the interstellar medium, and therefore represents the practical limit on how precisely gas can be kinematically associated with a specific Galactic location.
The derived central velocities are indicated in each figure panel.
We also estimated the total mass within the search regions, by integrating the gas hydrogen column density within the search region (dashed red circle). 
We notice that in several cases gas column densities exceeding $\approx 10^{22}$~cm$^{-2}$, which are typical of molecular cloud environments, are revealed.
These environments are favorable to the production of hadronic (neutral pion decay) gamma-rays, while inverse Compton scattering is a more likely mechanism for gamma-ray emission in low density environments.

Before proceeding, we comment on some of the associations.
The association with a 4FGL source of the cluster Mercer~23 is particularly interesting, as in this case the star cluster falls at the edge of the pointing accuracy ellipse of Fermi/LAT (95\% confidence) for the closest gamma-ray source 4FGL~J1929.8+1832 ($\Delta \vartheta =  0.1^{\circ}$).
Mercer~23 is also spatially associated with the unidentified LHAASO source 1LHAASO~J1928+1746u, which is classified as an ultra-high-energy gamma-ray source, whose spectrum extends beyond 100~TeV.
However, the center of the LHAASO source is located at a significantly larger  distance ($\Delta \vartheta = 0.55 ^\circ $)  than the nearest 4FGL source.
Finally, the cluster VVV~Cl036 is also interesting as it is located very close ($\Delta \vartheta = 0.09 ^\circ$) to the centroid of the extended Fermi/LAT source 4FGL~J1409.1-6121e, whose apparent radius is equal to 0.73~$^\circ$.
The entire region is characterized by the presence of dense molecular gas, that provides the target for CR interactions.

Another interesting feature for some clusters, as Bochum~14, is the large number of unidentified 4FGL sources found within the search region (6 in this case). 
Dedicated morphological and spectral analyses of such regions might be useful in order to determine whether the large number of unidentified sources might be in fact the signature of the presence of a large scale diffuse emission (as for example in the cases studied by \citealt{ge2022} and \citealt{Yang2020} for Car~OB1 and W~43, respectively).
We also note that Bochum~14 is located in a region of dense molecular gas, favoring an hadronic scenario for the possible origin of the gamma-ray emission.

We further notice that only unidentified 4FGL sources are found within the search region for Markarian~50 and VVV~Cl041. Dedicated analyses of Fermi/LAT data of these regions will thus be simplified by the lack of surrounding identified sources, although for Markarian~50 the presence of "c" sources would require careful treatment of the background. Moreover, for these regions, the low ambient gas density would favor leptonic over hadronic emission.

Finally, we note that the closest associated 4FGL unidentified sources all exhibit a soft gamma-ray spectrum, characterized by a power law with slope ranging from 2.4 to 2.7, except for Markarian 50 and VVV Cl036 with slope around 2.2 \citep{abdoll2020}. 
This is similar to what was found for a sample of very young clusters in \citet{peron2024}, but it should be kept in mind that steep spectra characterise a large fraction of 4FGL sources ($\sim$~40\% of them have slopes steeper than 2.4).

At this point, we should stress that Fig.~\ref{fig:maps} and Table~\ref{tab:associationsALL} indicate only spatial associations and not identifications (in other words, a number of the proposed associations might well be random coincidences). We present in Appendix. \ref{appendix:randomness} a statistical method that enables us to quantify the probability of random chance coincidence within the search regions. 
We limit our approach to the 4FGL, as most of the associations we found are with unidentified Fermi-LAT sources.
We find a hint ($\lesssim 3 ~\sigma$  confidence) for a spatial correlation between the star clusters listed in Table. \ref{tab:rankclusters} and unidentified Fermi-LAT sources. 

The search for spatial associations has been performed within a radius $\vartheta_s$ defined in Sec.~\ref{3.2}.
In order to explore the effects of the choice of the search radius on our results, we performed the search for spatial associations, and the simulations to determine the significance of a correlation for different values of $\vartheta_s$.
We tried radii of the search region around clusters in the range $R_s =$~10 to 50 pc, with a step of 10 pc, corresponding to $\vartheta_s \sim R_s/d$ and we found that the significance of the correlation increases to 4.7~$\sigma$ for a search region of $R_s =$~20~pc (more details are given in Appendix~\ref{appendix:randomness}.
This reinforces the hint for a correlation between star clusters and gamma-ray sources.

We also explore the random chance coincidences for individual cases and provide the results in Table.\ref{tab:statistics} an estimate of the randomness probability for all the clusters having associations with unidentified 4FGL sources. 
We stress that even when this probability is not small, further dedicated studies of the few regions shown in Fig.\ref{fig:maps} are encouraged. 
Indeed, dedicated gamma-ray analyses would help to better characterize the emission (morphology, spectrum, presence of diffuse emission, etc) and searches at other wavelengths could help to find counterparts of the unidentified gamma-ray sources.

Finally, we also searched for other sources within the search regions that could explain the origin of the gamma-ray emission from the unidentified gamma-ray sources.
In particular, we searched for pulsars (ATNF catalogue, \citealt{manche2005})
in the regions of interest that are not already firmly associated to gamma-ray sources. 
Both the pulsar itself \citep{smith2023} or its associated wind nebula \citep{Acero2013} can produce gamma rays.
For the pulsar wind nebulae scenario, we inspected the entire region of interest around each cluster, and we limited our search to pulsars characterised by a large ratio between the spin down luminosity $\dot{E}$ and the distance squared.
Following the work done in \citealt{Acero2013}, we set the limit to $\dot{E}/d^2 > 10^{35}$ erg/s/kpc$^2$, as these objects are expected to power a nebula which is detectable in gamma rays. 
We find that no pulsar satisfies this criterion.
On the other hand, gamma rays produced by the pulsar itself can be detectable even for small values of $\dot{E}/d^2$.
\citep{smith2023} performed a search for pulsars which are cospatial with 4FGL sources and have a sufficiently large $\dot{E}/d^2$ to power the gamma-ray emission.
Only one unidentified gamma-ray source that we found within the search regions around clusters is present in the list by \citet{smith2023}. It is 4FGL~J1631.7-4826c, in the search region of VVV Cl073, cospatial with the pulsar J1632-4818.

Supernova remnants are also present within the search regions, but in most cases they are identified with gamma-ray sources (see Table~\ref{tab:associationsALL}). We searched, within the search regions, for supernova remnants from the Green catalogue \citep{green2025} which are not associated to any known gamma-ray source (they could be isolated or embedded in the corresponding clusters) and we found four of them: G006.5-00.4 (for Bochum~14), G298.5-00.3 (Dragonfish), G053.4+00.0 (Mercer~23), and G311.5-00.3 (VVV~Cl036).
An additional object, G312.4-00.4, found within the search region around VVV~Cl036, is now believed to be associated with part of 4FGL~J1409.1-6121e (which has been resolved into three distinct sources \citep{chambery2023}).

\section{Spatial associations of isolated WR stars and gamma-ray sources}
\label{sec:rankWR}

In Sec.~\ref{sec:rankclusters} we proposed a method to rank star clusters and associations according to the ratio between the mechanical power injected by member stars and the distance squared $L_w/\langle d \rangle ^2$.
The method is based on counting the number of WR stars $N_{WR}$ hosted by a given cluster/associations, as is motivated by the fact that a proportionality is expected between $N_{WR}$ and $L_w$.
Moreover, in Sec.~\ref{sec:power} we showed that the mechanical power injected by a single WR stellar wind can compete with the total wind power injected by an entire star cluster.
We concluded then that even a single WR star might power a gamma-ray source detectable by current instruments.
Searching for spatial associations between gamma-ray sources and individual and powerful WR stars is of particular relevance as more than 60\% of those stars are indeed isolated \citep{rate2020}. 
For this reason, in the remaining of this Section we will report on the results of such search.

To do so, we start with the same sample of 108 WR stars that we considered in Sec.~\ref{sec:power}.
For these stars, accurate estimates of $L_w$ and $d$ have been provided by \citet{sander2019} and \citet{hamann2019}.
Following a similar reasoning as in Sec.~\ref{sec:rankclusters}, we select individual WR stars in our sample for which $L_w/d^2 \ge 10^{37}$~erg/s/kpc$^2$. 
This roughly corresponds to the minimum $L_w/d^2$ of the star clusters considered in \citealt{aharon2019} and associated with gamma-ray emission (see their Table~1). 
We found that 16 WR stars from our sample satisfy this condition \citep[for the full list see][]{inventar2025}.
We further restrict our sample by eliminating all stars that, according to Crowther's catalogue of WR stars, belong to a cluster \citep{roscro2015}.
More precisely, following Crowther's nomenclature, we remove from the list stars for which the association with a cluster is firm or likely, and we keep stars for which no association is reported, those for which an association with a cluster is unlikely, and those for which an association was proposed but then questioned.
This reduces the list to 10 stars, which are listed in Table~\ref{tab:WRs} and ordered according to the value of $L_w/d^2$.
Note that except for WR~147, 90, and 52, all other stars in Table~\ref{tab:WRs} are located in projection against star clusters, but most likely they do not belong to them (see \citealt{inventar2025} for more details).

\begin{figure*}[ht]
    \centering
    \begin{subfigure}[t]{0.34\textwidth}
        \centering
        \includegraphics[width=\linewidth]{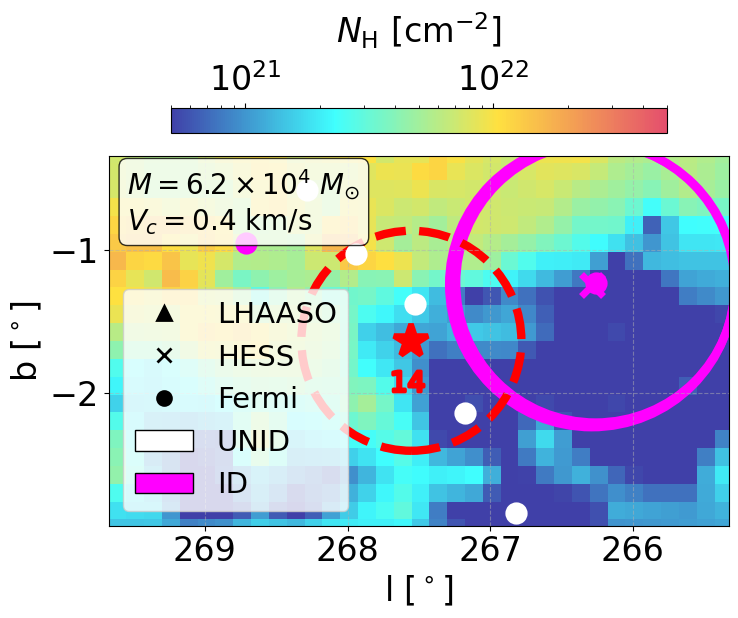}
    \end{subfigure}
    \hspace{-2mm}
    \begin{subfigure}[t]{0.3\textwidth}
        \centering
        \includegraphics[width=\linewidth]{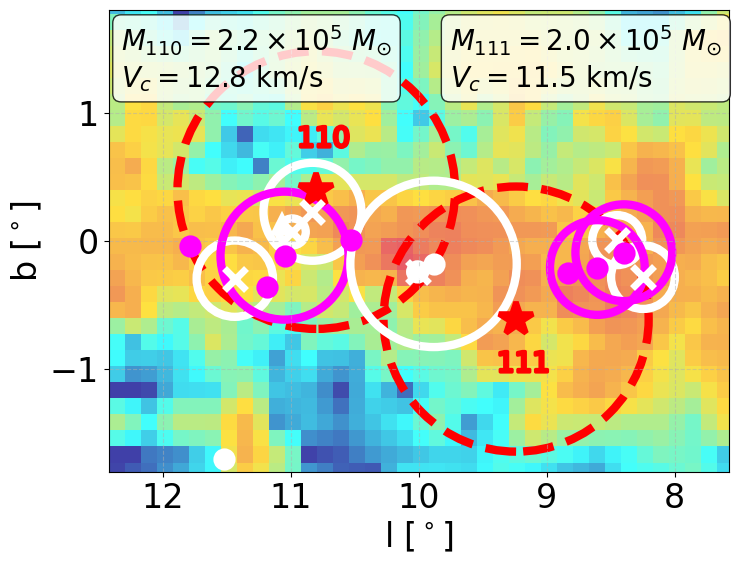}
    \end{subfigure}
    \hspace{-2mm}
    \begin{subfigure}[t]{0.313\textwidth}
        \centering
        \includegraphics[width=\linewidth]{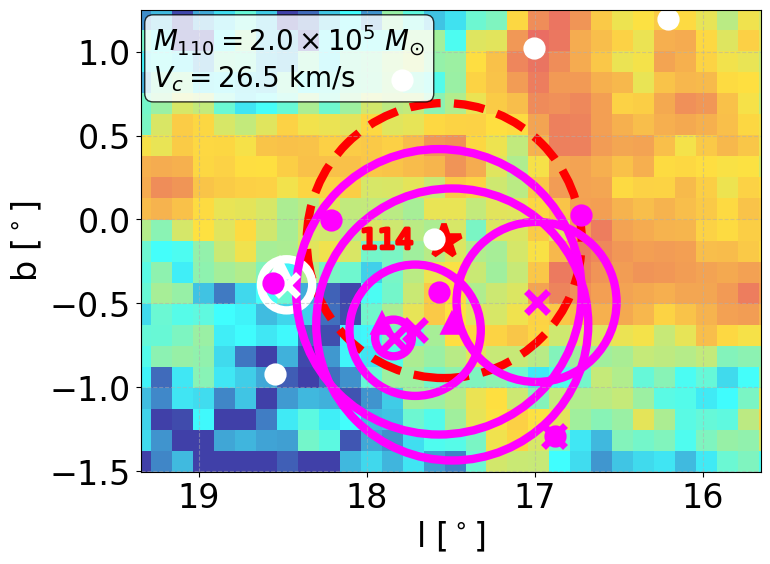}
    \end{subfigure}
    
    \caption{Position (magenta/white points) of gamma-ray sources found in the vicinity of the WR stars from Table~\ref{tab:WRs} for which a spatial association with unidentified gamma-ray sources has been found. If the source is not point-like, its extension is indicated with a circle. The position of the WR star is indicated by a star and the red dashed circle shows the search region (see text for more details). The background color indicates the hydrogen gas column density, integrated in a range of $\pm 10 \rm ~ km/s$ around the velocity $V_c$ corresponding to the distance of the WR star. The total hydrogen mass in the search region is also indicated.}
    \label{fig:mapsWRs}
\end{figure*}

\begin{table}[H]
\caption{\centering Isolated WR stars from our sample ranked according to $L_w/d^2$ and with $L_w/d^2 \ge 10^{37}$~erg/s/kpc$^2$.}
\begin{center}
\begin{tabular}{| c | c | c | c |}
\hline
WR & $L_w/10^{37}$ & $d$ & $L_w/d^2/10^{37}$ \\
& [erg/s] & [kpc] & [erg/s/kpc$^2$] \\
\hline
110 & 10.6 & 1.58 & 4.2 \\
\hline
147 & 5.03 & 1.20 & 3.5 \\
\hline
90 & 3.44 & 1.15 & 2.6 \\
\hline
114 & 10.0 & 2.10 & 2.3 \\
\hline
52 & 5.86 & 1.75 & 1.9 \\
\hline
15 & 16.4 & 2.98 & 1.9 \\
\hline
111 & 4.18 & 1.66 & 1.5 \\
\hline
136 & 5.12 & 1.91 & 1.4 \\
\hline
14 & 6.22 & 2.23 & 1.3 \\
\hline
134 & 3.65 & 1.74 & 1.2 \\
\hline
\end{tabular}
\tablefoot{The star identifier (first column) is from Crowther's catalogue.}
\label{tab:WRs}
\end{center}
\end{table}

In the following, as done for clusters in Sec.~\ref{sec:rankclusters}, we search for spatial associations between the WR stars in Table~\ref{tab:WRs} and gamma-ray sources from the 4FGL, the HGPS, and the First LHAASO Catalog.
Similarly to what done above, we use a search region centered on the star and of radius $\vartheta_s \equiv \vartheta_{30}$ (defined in Sec.~\ref{sec:rankclusters}). 
Unidentified gamma-ray sources are found within the search regions for 5 of the stars considered: WR~110, 114, 111, 14 and 147. 
However, we will not further consider here WR~147, because it is associated with the source LHAASO J2032+4102, which is believed to belong to the complex Cyg~OB2 region \citep{lhaaso2024}, which we already discussed above (Sec.~\ref{3.2}). 
For the 4 other stars, results are shown in Fig.~\ref{fig:mapsWRs}, which is formatted exactly as Fig.~\ref{fig:maps}.
The list of all sources located within the search region can be found in Table~\ref{tab:associationsALLWR}.
Also in this case, the possibility that the associations are random spatial coincidences cannot be ruled out a priori. 
We conducted the same statistical tests than was performed for clusters in Appendix \ref{appendix:randomness}. 
We do not find evidence for a spatial correlations between the WR stars considered in Table~\ref{tab:WRs} and unidentified 4FGL sources.
As done for clusters, We also explored the random chance coincidences for individual cases and provide the results in Table.\ref{tab:statistics}.

Also in this case, dedicated analyses/observations of these very few regions are highly encouraged in order to confirm these associations or rule them out.
Dense molecular gas is found in all cases, with the largest column densities found within the search regions of WR~110 and WR~111.

In the following we discuss in detail the 4 stars for which spatial associations have been found.

\subsection{WR~110: escape from the acceleration region?}
\label{sec:WR110}

Among the WR stars in Table~\ref{tab:WRs} that do not belong to any cluster/association, WR~110 is the one characterized by the largest value of the ratio $L_w/d^2$ (but \citealt{blanco2024} reported an estimate of the wind power smaller by a factor of a few).
The star is spatially associated to several unidentified sources, the closest of them being the TeV source HESS~J1809-193.
Therefore, in order to check if such an association is plausible, the question arises whether particles can be accelerated at multi-TeV energies at a stellar WTS.

To answer that we follow a reasoning very close to that described by \citet{harer2025}. 
We consider a point-like injection of mechanical energy by a single Wolf-Rayet star in the form of a spherical wind of constant speed $u$ and stationary mass loss rate $\dot{M}$ at a rate $L_w = (1/2) \dot{M} u^2$. Thus, the wind density profile scales as $\varrho = \dot{M}/4 \pi u r^2$, where $r$ is the distance from the injection site, and the wind will be decelerated at a WTS located at $r = R$.
A necessary condition for the formation of the shock is that the flow of matter across it has to be super-Alfvenic. Considering a quasi-stationary WTS this conditions reads $u > B/\sqrt{4 \pi \varrho}$, and can be inverted to get an upper limit for the strength of the magnetic field at the shock: $B < (2 L_w/u R^2)^{1/2}$.
At this point, we can use the Hillas criterion to obtain the most optimistic estimate of the maximum energy achievable by protons accelerated within the system \citep{hillas1984,hillas2005}:
$E_{max} \approx (e/c) u B R$,
where $e$ is the elementary charge.
This can be rewritten as:
\begin{equation}
\label{eq:Emax}
E_{max} < \xi \frac{e}{c} \sqrt{2 L_w u} \sim 2.4 ~\xi \left( \frac{L_w}{10^{38} {\rm erg/s}} \right)^{1/2} \left( \frac{u}{3000~{\rm km/s}} \right)^{1/2} \rm PeV
\end{equation}
where power and wind speed have been normalised to the extreme values appropriate to describe the most powerful winds of WR stars \citep[][]{sander2019,hamann2019}.
Remarkably, this upper bound for $E_{max}$ does not depend on the WTS radius $R$.
Eq.~\ref{eq:Emax} shows that, under most optimistic conditions, protons from single WTS might be accelerated up to PeV energies in the most optimistic scenario. 
However, as noticed by \citet{harer2025}, the Hillas criterion provides an overly optimistic estimate for $E_{max}$.
If more realistic models are used to describe the structure of the magnetic field in the wind \citep[e.g. those in][]{usomel1992}, the estimate of the value of $E_{max}$ is significantly reduced.
That is why we introduced the factor $\xi$ which should typically be set equal to $\approx 0.1-0.2$ \citep{harer2025}.
The spectrum of accelerated protons may extend then up to hundreds of TeV, and therefore the related gamma-ray emission might well extend up to the multi-TeV domain.

An association between WR~110 and the extended TeV source HESS~J1809-193 seems possible.

The star is also located in the vicinity of the extended (flat disk of radius $0.5^{\circ}$) Fermi source 4FGL~J1810.3-1925e (which is identified as pulsar in the 4FGL) and of the ultra-high energy gamma-ray source 1LHAASO J1809-1918u. 
The region around these sources is quite complex, hosting the powerful pulsars PSR J1811-1925 ($\dot{E}/d^2=2.6 \times 10^{35} \rm{~erg/s/kpc^2}$, \citealt{manche2005}) and PSR J1809-1917 ($\dot{E}/d^2=1.3 \times 10^{35} \rm{~erg/s/kpc^2}$),
several supernova remnants from Green's catalogue
(G011.1+00.1, G011.0-00.0, G011.4-00.1, and G011.2-00.3),
and molecular clouds. 
Recent studies have tried to ascribe the origin of the gamma-ray emission to one of these objects \citep{Castelletti_J1809,hess_J1809-193,Hawc_J1809} but no definitive conclusion has been drawn. 
Remarkably, the $10^{38}$~erg/s of mechanical energy provided by the WR wind might constitute a substantial contribution to the overall gamma-ray emission.

\begin{figure}
   \centering
   \includegraphics[width=0.45\textwidth]{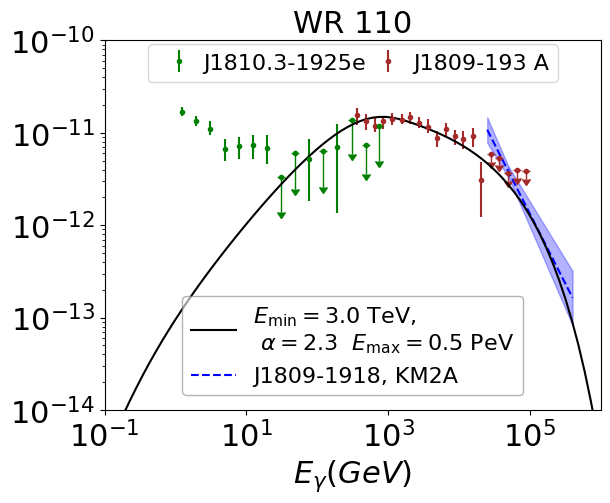}
   \caption{Spectrum of the gamma-ray sources tentatively associated with WR~110. The solid line shows the predicted $\pi^0$-decay flux (see text). Green, brown, blue data points refer to Fermi/LAT, HESS, LHAASO (KM2A) observations, respectively.}
    \label{fig:WR110}
 \end{figure}

As the region hosting WR~110 is very crowded (see Table~\ref{tab:associationsALLWR} the long list of sources within the search region), we show in Fig.~\ref{fig:WR110} the spectra of all of the sources mentioned above, i.e. 4FGL~J1810.3-1925e, LHAASO~J1809-1918, and HESS~J1809-193.
For the latter, in a dedicated paper the H.E.S.S. collaboration (2023) identified two spatial components (named A and B) contributing to the TeV emission, and reanalysed Fermi/LAT data for 4FGL~J1810.3-1925e. 
In Fig.~\ref{fig:WR110} we show this reanalysis plus the H.E.S.S. data for the component A, for which $\Delta \vartheta / \vartheta_s \sim 0.49$.
The H.E.S.S. Collaboration
also noticed that due to a significant spectral difference it is not straightforward to interpret simultaneously the GeV and TeV spectra from the region. 
As WR~110 lays towards the edge of component A of HESS~J1809-193, and given the presence of molecular clouds in the region, the emission could be in principle associated to CRs that escaped from the WTS and now interact with some dense gas.
To test the plausibility of such a scenario we assume that CRs are injected at the WTS and that only those of highest energy (those diffusing faster) reached the dense region.
The curve in Fig.~\ref{fig:WR110} shows the gamma-ray emission from proton-proton interactions \citep{kafexh2014} for a CR spectrum (per unit volume) in the cloud which is $n(E) = n_0 E^{-2.3} \exp{(-E/E_{max})}$ above a given particle energy $E_{min}$~
and zero otherwise \citep[cfr.][]{Gabici2009}.
To fit data we set $E_{min} = 3$~TeV and $E_{max} = 500$~TeV and this allows us to fix a value of $n_0$ for an assumed cloud mass $M_{cl}$ (because for a fixed gamma-ray flux $n_0 \propto M_{cl}^{-1}$).

If CRs have been injected during the WR phase that started $\tau_{WR}$ years ago, and diffused then isotropically in the surrounding medium, then the minimum energy $E_{min}$ can be obtained imposing $R \approx \sqrt{D(E_{min}) \tau_{WR}}$, where $D = D_0 (E/{\rm GeV})^{\delta}$ is the CR spatial diffusion coefficient (here we fix $\delta = 1/3$ and R the distance between the WR star and the target gas for CR interactions.
This allows us to infer $D_0 \approx 10^{26} (R/20~{\rm pc})^2 (\tau_{WR}/10^5 {\rm yr})^{-1} $~cm$^2$/s, which is significantly smaller than the typical Galactic diffusion coefficient.
This may indeed be the case, as small diffusion coefficient have been inferred from observations of the vicinities of powerful CR accelerators \citep[see e.g.][]{gabici2010}.
For energies larger than $E_{min}$ the steady state solution of the transport equation is recovered and one can write $n(E) = Q(E)/4 \pi D(E) R$.
The total energy injected in form of CRs since the beginning of the WR phase can be then computed as $E_{tot}=\tau_{WR} \int_{\rm 1 GeV}^{\rm 1PeV}EQ(E)dE $,
which gives $E_{tot}\sim 8 \times 10^{48} (M/10^5 \rm M_{\odot})^{-1} (R/20~{\rm pc})^3$~erg.

This number can be compared to the mechanical energy injected by the WR star over a time $\tau_{WR}$, which is $\sim 3 \times 10^{50} (\tau_{WR}/10^5 {\rm yr})$~erg.
The emission can be reproduced if a fraction $\approx 0.03 ~(M/10^5 \rm M_{\odot})^{-1} (R/20~{\rm pc})^3 (\tau_{WR}/10^5 {\rm yr})^{-1}$ of the wind mechanical energy is converted into CRs.
We note that in order to explain the highest energy part of the spectrum $E_{max}$ had to be set equal to 500~TeV, which is close to the maximum allowed value computed above (Eq.~\ref{eq:Emax}).
For the estimated physical parameters of WR~110 ($L_w \sim 1.1 \times 10^{38}$~erg/s and $u_w \sim 2300$~km/s), a maximum energy of 500 TeV requires $\xi \approx 0.2$.

\subsection{WR~111 and 114: gamma rays from a portion of the wind termination shock?}

\begin{figure*}
   \centering
   \includegraphics[width=0.43\textwidth]{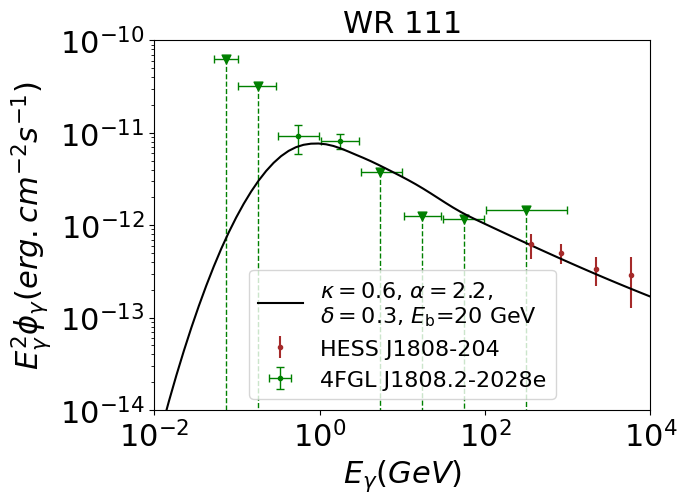}
\includegraphics[width=0.43\textwidth]{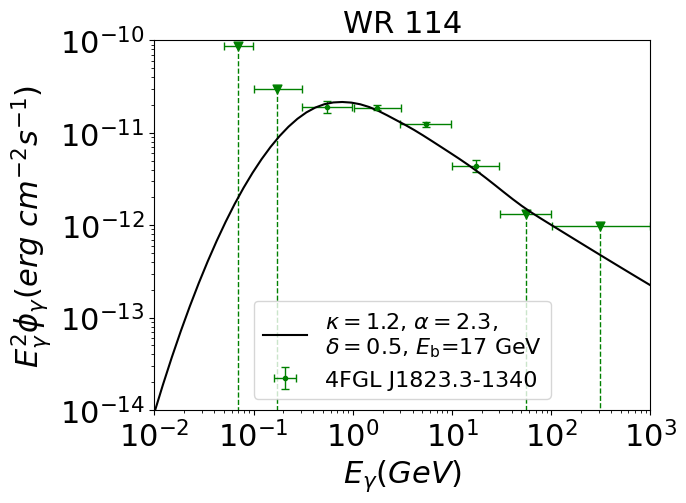}
   \caption{Spectrum of the gamma-ray sources tentatively associated with WR~111 (left) and WR~114 (right). The solid line shows the predicted $\pi^0$-decay flux (see text). Green and brown data points refer to Fermi/LAT and HESS observations, respectively.}
    \label{fig:WR111}
 \end{figure*}

The GeV source 4FGL~J1808.2-2028e and its TeV counterpart HESS~J1808-204 are the closest unidentified gamma-ray sources to WR~111. 
Both sources are extended, the former fitted with a uniform disk model of radius 0.65$^{\circ}$, the latter by a 2D Gaussian of 39\% containment radius of $0.06^{\circ}$.
Also in this case the WR star is located in a complex region, characterized by the presence of molecular clouds (more than $10^5 M_{\odot}$ of molecular hydrogen are found within the search region, see Fig.~\ref{fig:mapsWRs}), three supernova remnants about which little is known\footnote{They are SNR G009.7-00.0, G009.8+00.6, and G009.9-00.8 \citep{green2025}.}, and the star cluster SGR1806-20 (which was also proposed to power the H.ES.S. source, see \citealt{Yeung2016} and Appendix~\ref{sec:appendix}).
On the other hand, no powerful pulsar is found within the search region.
Keeping in mind that many possibilities exist to explain the gamma-ray emission, it seems appropriate to check also a scenario in which the emission is powered by the stellar wind blown by WR~111. 

\color{black}

Here, we propose a possible interpretation of the gamma-ray spectrum in terms of hadronic (proton-proton) interactions \citep{kafexh2014}. Indeed, a hadronic scenario seems favored over a leptonic one, due to the presence of dense gas in the region (see Fig.~\ref{fig:mapsWRs}) and to the steepness of the gamma-ray spectrum, that would require an exceedingly steep spectrum of CR electrons \citep[see][]{peron2024}.
We assume that a fraction $\eta$ of the power of the wind blown by WR~111 is converted into CR nuclei (mostly protons).
These energetic protons interact with ambient matter of density $n$ and produce neutral pions ($p + p \rightarrow p + p + \pi^0)$ that in turn decay into two gamma rays \citep{kafexh2014}.
Protons lose energy due to $\pi^0$ production on a time scale which is almost energy independent and equal to $\tau_{pp \rightarrow \pi^0} = 3 \tau_{pp} = 3/ (n \sigma_{pp} K c) \approx 180 ~(n/{\rm cm}^{-3})^{-1}$~Myr, with $\sigma_{pp}$ and K the cross section and inelasticity of p-p interactions. 
As most of the energy channeled in gamma-rays falls well within the 100 MeV-100 GeV photon energy range, we can write the approximate expression:
\begin{equation}
\label{eq:eq}
F_{\gamma} \approx \frac{\eta L_w}{4 \pi d^2} \left( \frac{\tau_{res}}{\tau_{pp \rightarrow \pi^0}} \right) 
\end{equation}

where $\tau_{res}$ (the residence time of CRs within the considered region) is now at most the duration of the WR phase, $\tau_{WR}$.
As $L_w$ and $d$ can be taken from \citet{sander2019},
we can equate the expression above to the observed gamma-ray to show how gamma-ray observations allow us to constrain the quantity $\kappa$ introduced in the beginning of Section \ref{sec:rankclusters} and that we set here equal to $\kappa \equiv (\eta/0.1) (\tau_{res}/10^5 {\rm yr}) (n/100~{\rm cm^{-3}})$, where we normalized the ambient density to the typical value of molecular clouds).

In fact, the morphology of the gamma-ray emission will depend on both the location of the acceleration site (a natural candidate being the spherical stellar WTS, whose expected radius is of the order of few tens of parsecs) and the distribution of target material in the region (which can be quite inhomogeneous on such spatial scales).
The fact that the gamma-ray emission comes only from a fraction of the search region may be interpreted as the fact that a concentration of target gas is present there.
This means that not all the mechanical power of the wind is used to power the gamma-ray emission, but only a fraction of it.
To make our estimate more realistic we introduce then a geometrical factor $f$, and we redefine $\kappa$ as $(\eta/0.1) (\tau_{res}/10^5 {\rm yr}) (n/100~{\rm cm^{-3}}) (f/0.1)$.
In other words, introducing $f$ is equivalent to using $f L_w$ instead of $L_w$, and $f$ represents the fraction of the solid angle around the WR star which is covered by dense gas.

\color{black}

In Fig.~\ref{fig:WR111} we show a model curve to explain the gamma-ray data for WR~111.
We assume that the WTS of WR~111 accelerates CR protons at a constant rate $Q(E) \propto E^{-\alpha}$, defined in such a way that $\eta L_w = \int_{1 \rm GeV}^{1 \rm PeV} {\rm d}E Q(E) E$.
We allow particles of energy exceeding $E_b$ to leave the system in a characteristic time $\tau_{esc}(E)$.
If the WR phase started $\tau_{WR}$ years ago and the escape is diffusive, then $\tau_{esc} \sim L^2/D(E) \propto E^{-\delta}$, where $L$ is the size of the system, $D$ the diffusion coefficient,
and $E_b$ can be derived imposing $\tau_{esc}(E) = \tau_{WR}$.
To mimic that, we assume that the spectrum of protons in the region is $Q(E) \times \tau_{WR} \propto E^{-\alpha}$ for $E < E_b$ and $Q(E) \times \tau_{esc}(E) \propto E^{-\alpha-\delta}$ for $E > E_b$.
It is easy to show that the emission is proportional to the parameter $\kappa$ defined above.
We then compute the gamma-ray emission due to proton-proton interactions following \citep{kafexh2014}.
The adopted physical parameters are reported in the figure inset.
We notice that data are well explained for a very reasonable value $\kappa = 0.6$.

We now consider the case of WR~114.
This WR star is located at 0.06$^{\circ}$ only from the centroid of the unidentified Fermi source 4FGL~J1823.3-1340.
This source is detected up to photon energies exceeding 10~GeV \citep{ajello2017}
and has no very-high-energy counterpart according to TeVCat \citep{wakhor2008}.
Its 100~MeV-100~GeV flux is $F_{\gamma} \sim 7.5 \times 10^{-11}$~erg/cm$^2$/s, and the spectrum is fitted by a log-parabola with photon index steepening from $\alpha \sim 2.1$ to $\sim 2.6$ for photon energies going from 1 to 10 GeV.
It is classified as unidentified on 4FGL, and a literature search reported only a tentative association with the globular cluster Mercer 5 \citep{hui2020}. 
We  also note that a powerful pulsar (PSR J1826-1334, $\dot{E}/d^2=1.7\times 10^{35} \rm{~erg/s/kpc^2}$) is present at the edge of the search region, i.e. $\sim$30~pc away from the star)
as well as some supernova remnants from the Green catalogue. They are G017.4-00.1, G017.0-00.0, G018.1-00.1 (associated with 4FGL J1824.1-1304, see Tabel~\ref{tab:associationsALLWR}).

We repeated the same procedure as for WR~111, and we show in figure~\ref{fig:WR111} (right panel) that a fit to data can be obtained adopting $\kappa = 1.2$.
However, this fit might be more problematic than that obtained for WR~111 because the associated gamma-ray source, 4FGL~J1823.3-1340, is point-like, and therefore a small value of the geometrical factor $f$ is probably more appropriate.
Moreover, the estimate of $\kappa$ may increase by a factor of few if a smaller estimate for the wind power is adopted, such as that from \citet{saha2023}, based on radio observations.

\subsection{WR~14: leptonic emission?}

\color{black}

The unidentified GeV source closest to WR~14 is 4FGL~J0856-4724c, a weak source labeled as confused in the 4FGL (the letter c at the end of its name indicates that the source could be in fact confused with diffuse emission), which has a flux equal to $F_{\gamma} \sim 1.9 \times 10^{-12}$~erg/cm$^2$/s. 
Two more unidentified 4FGL sources are present within the search region, 4FGL~J0851.2-4737 and 4FGL~J0859.2-4729, but the latter coincides and has been associated with the HII region RCW38 \citep{peron2024}. No powerful pulsars, nor SNRs are present in the region.

We note that the region surrounding the WR star is characterized by gas column densities smaller than $\sim 10^{22}$~cm$^{-2}$.
Column densities are particularly small in correspondence with the two unidentified sources 4FGL~J0856-4724c and 4FGL~J0851.2-4737, making and hadronic origin of emission less likely.

\begin{figure}
   \centering
   \includegraphics[width=0.43\textwidth]{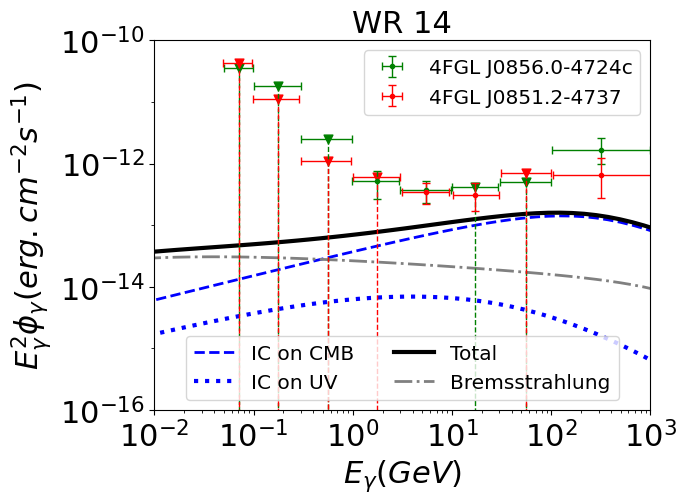}
   \caption{Spectrum of the unidentified gamma-ray sources found in the search region around WR~14 (green and red data points). The solid lines show a benchmark estimate for the leptonic emission expected from the region (see text for more details).}
    \label{fig:WR14}
 \end{figure}

For this reason, we investigate here the possible role of leptonic (inverse Compton and non-thermal Bremsstrahlung) emission from the vicinity of the star.
We notice that the inverse Compton emission is expected to be extended (as soft target photons in the ambient radiation) and trace the acceleration region (for example, the WTS), while the non-thermal Bremsstrahlung component will follow the ambient gas distribution.
The curves in Fig.~\ref{fig:WR14}, done with the GAMERA code \citep{Hahn2022}, show the leptonic emission (including Klein-Nishina effects) expected from the region if 0.1\% of the wind mechanical energy is converted into relativistic electrons of spectrum $\propto E^{-2.2}$ extending up to a particle energy of $30$~TeV.
Such an acceleration efficiency seems appropriate for electrons accelerated at strong shocks \citep[see e.g.][]{cristofari2013}.
Injection is assumed to be continuous over $10^5$~yr and we accounted for energy losses due to synchrotron (ambient magnetic field equal to 3~$\mu$G), inverse Compton scattering (on the cosmic microwave background radiation and on the UV photons from the star), non-thermal Bremsstrahlung and ionization losses (ambient density set equal to 1~cm$^{-3}$).
The radiation energy density from the star is estimated as $U_{ph}= L/4 \pi R^2 c$ where $L$ is the star bolometric luminosity and $R$ is a reference distance that we set equal to 10~pc. The spectrum of the radiation is a gray-body of temperature $T$.
Both $L$ and $T$ have been taken from \citet{sander2019}.

Our benchmark prediction is shown in Fig.~\ref{fig:WR14} together to the spectra of the unidentified 4FGL sources found within the search region.
The model is of course not intended to provide a fit to the data, but rather as an illustrative example demonstrating that the expected extended emission from inverse Compton scattering lies at a level potentially accessible to more sophisticated analyses or deeper observations.

\color{black}

\section{Discussion and conclusions}
\label{sec:conclusion}

In this work we investigated the role of WR stellar WTSs as particle accelerators, and we discussed the possible related radiative signatures in the gamma-ray domain.
The study is motivated by the large mechanical wind power that characterizes these objects.
We considered both the case of WR stars belonging to stellar clusters and associations, and that of isolated WR stars.

Our study is driven by the fact that the mechanical power of the wind of a single WR star can be comparable to the global wind power of the most massive Galactic stellar clusters.
This led us to propose a method to estimate the global mechanical power of stellar systems based on counting the number $N_{WR}$ of WR stars that they host, a quantity that is expected to increase as the overall wind power $L_w$ of the system.
We then ranked stellar clusters and associations according to the value $N_{WR}/d^2$, where $d$ is the distance to the system.
We did that because a large value of $N_{WR}/d^2$ is a necessary condition to have large gamma-ray fluxes.
Our ranking is quite similar to the one proposed by \citet{celli2024} for star clusters, but it has the advantage of being very straightforward and computationally inexpensive, and applicable also to stellar associations.
On the other hand, our ranking does not include clusters which are too young to host WR stars, and therefore our approach and Celli's can be considered, with this respect, complementary. 

Only a handful of star clusters/associations are currently believed to be associated with gamma-ray sources, and they are not necessarily the ones characterized by the highest values of $L_w/d^2$.
For this reason, we suggest that environmental effects must play an equally important role in shaping the gamma-ray emission of these objects.
In an attempt to increase the number of such objects detected in gamma-rays, we searched for new spatial associations between clusters hosting WR stars and unidentified gamma-ray sources from the 4FGL, the HGPS and the First LHAASO Catalog.
We found that 11 objects (over 29 considered) may be potentially associated with Fermi/LAT source, and 2 of them also with ultra high energy gamma-ray sources detected by LHAASO.

We then developed a method to assess the probability that the spatial coincidence are random, and found a hint ($\lesssim$ 3~$\sigma$ confidence) for a correlation between star clusters hosting WR stars and unidentified 4FGL sources.
Dedicated analyses of existing data or deeper observations with gamma-ray instruments will help in characterizing the emission (e.g. morphology, spectra...). On the other side, a search for more possible counterparts of the unidentified gamma-ray sources will help to confirm or rule out the proposed associations.

Finally, we also searched for spatial association between isolated WR stars and gamma-ray sources.
Also in this case the search is motivated by the fact that the power of the WTS of a single star can be comparable to that of entire stellar clusters.
We considered a large sample of 108 WR stars for which the wind power is estimated based on an accurate modeling of stellar spectra.
We then repeat the same procedure performed for stellar clusters/associations: we ranked WR stars according to $L_w/d^2$ and we searched for spatial associations between the top ranked stars and gamma-ray sources.
We found four spatial associations between isolated WR stars (WR~110, 114, 111, and 14) and gamma-ray sources, and we model the gamma-ray emission in terms of interactions between CRs accelerated at the WTS and ambient gas or photon fields.
In this case, we do not find statistical evidence for a significant spatial correlation between the considered WR stars  (Table~\ref{tab:WRs}) and 4FGL sources.

Similarly to clusters, further studies on the proposed associations are encouraged. If confirmed, these associations will establish WR stars as a new class of gamma-ray emitter.

We conclude by noticing that we considered a large but highly incomplete sample of WR stars (about 15\% of all known Galactic WR stars, which are in turn probably a half of all existing Galactic WR stars).
It would be interesting to extend the study to all known WR stars, as well as to O stars  \citep[even though they tend to be more clustered than WR stars][]{maiz2013}.
This would require to extend the estimates of wind powers to a sample of star much larger than that considered here.

\begin{acknowledgements}
We enjoyed many interesting discussions with the participants to TOSCA: Topical Overview on Star Cluster Astrophysics, held in Siena on October 2024.
We thank P. Crowther, A. Sander, P. Martin, and R. Terrier for useful comments and advices.
We also thank the editor, S. Campana, and the referee for constructive feedback.
\end{acknowledgements}

\bibliographystyle{aa} 
\bibliography{bib.bib} 

\begin{appendix}

\section{Tentative associations between clusters and gamma-ray sources from Table~\ref{tab:rankclusters}}
\label{sec:appendix}

Here we briefly review the clusters that, according to our literature search, have been tentatively associated with gamma-ray emission (labeled as ``Tentative'' in the last column of Table~\ref{tab:rankclusters}). 

\textit{Galactic center:} The diffuse TeV emission observed by H.E.S.S. from the inner $\approx 100$~pc of the Galaxy has been interpreted as the result of hadronic interactions of CRs accelerated at an object located very close to the Galactic center \citep{hessco2016}.
It has been suggested that the accelerators of these CRs could be the three clusters (Arches, Quintuplet, and Nuclear) located very close to the Galactic center \citep{aharon2019}.
However, other sources have also been proposed to explain the diffuse gamma-ray emission (most notably, the supermassive back hole at the center of the Galaxy).

\textit{Car OB1:} This OB association is very loose ($\theta_{max} \sim 3 ^\circ$), and it has been tentatively associated to a diffuse GeV emission by \cite{ge2022}.
However, this is a very crowded and complex region, and it is not easy to draw firm conclusions.
Following the criterion defined in Sec.~\ref{sec:rankclusters},
we found 22 4FGL sources (8 unidentified and 14 identified, including the colliding wind binary $\eta$-Carinae) that satisfy the condition $\Delta \theta /\theta_s <1$. 
No TeV source satisfies the condition $\Delta \theta /\theta_s <1$.

We notice that the star WR~25 ($L_w/d^2 \sim 1.2 \times 10^{37}$~erg/s/kpc$^2$) is a member of a binary system and of the star cluster Trumpler~16, which is in turn part of the Car~OB1 association \citep{inventar2025}, and it may contribute to power the gamma-ray emission from the region (even though upper limits of the GeV flux from the position of the star have been reported in \citealt{Marti_devesa2021}).
The closest unidentified source to WR~25 is 4FGL~J1046.7-6010 ($\Delta\theta/\theta_s=0.64$).

\textit{Mercer~20:} Diffuse gamma-ray emission has been detected in the surrounding of this cluster \citep{Sun2022}, with an angular extension of $\sim 0.7^\circ$ and a centroid offset by $\sim 0.4^\circ$ from the cluster position. We found four unidentified 4FGL sources in our search region, the closest being 4FGL 1912.7+0957 and with $\Delta\theta/\theta_s =0.1$ only. We also find 1 identified Fermi source classified as pulsar, together with an HESS and a LHAASO identified source, classified as SNR. The presence of such objects makes it difficult to interpret this diffuse emission, but the cluster could play an important role in injecting CRs in the region.

\textit{Danks 1 and 2:} Also in this case, diffuse GeV emission has been tentatively associated with these objects \cite{liu2024}.
We could not find in the literature any detailed investigation on possible alternative sources for the origin of the gamma-ray emission. 
We found 3 (2) unidentified 4FGL sources within a distance $\vartheta_s$ from Danks~1 (Danks~2), and no TeV source.
We also searched for supernova remnants \citep{green2025} and pulsars \citep{manche2005} in the region, but could not find any.
This reinforces the hypothesis that the two star clusters might be at the origin of the diffuse gamma-ray emission.

\textit{Berkeley~87}: This cluster has been tentatively associated with two EGRET and one MILAGRO sources by \citet{mancha1996} and \citet{bednar2007}.
We found two 4FGL sources (J2026.5+3718c, J2022.6+3716c) and two LHAASO catalogue sources (J2020+3638, J2020+3649u) within the search region around the cluster.
However, one identified source is also found both in 4FGL (J2021.1+3651) and LHAASO catalog (J2018+3643). 
These two sources are associated and correspond to the pulsar wind nebula powered by PSR~J2021+3651.

Remarkably, the WR star in our sample which is characterised by the largest value of $L_w/d^2$, WR~142, belongs to Berkeley~87 \citep{inventar2025}.
With $L_w/d^2 \sim 4.6 \times 10^{37}$~erg/s/kpc$^2$, it may certainly power a gamma-ray source.
The closest unidentified sources to WR~142 are 4FGL~J2022.6+3716c ($\Delta\theta/\theta_s=0.19$) and 1LHAASO~J2020+3649u ($\Delta\theta/\theta_s=0.56$).

Interestingly, \citet{Romero1999} proposed that this star could be a gamma-ray emitter based on the association with an EGRET source.

\textit{Cl 1813-178:} The source HESS J1813-178 and a counterpart found in Fermi/LAT data have been tentatively associated to this cluster by \citet{guo2024}, who also noted that the region is very complex and rich of potential CR accelerators that may also explain the gamma-ray emission (most notably the supernova remnant G12.820.02 and the pulsar wind nebula powered by PSR J1813-1749).
We found that the unidentified source 1LHAASO~J1814-1719 is very close to the cluster ($\Delta \vartheta/\vartheta_s$=0.09).

\textit{Mercer 81:} Due to its proximity, it was tentatively associated to the source HESS~1640–465 \citep{davies2012}, which in the HGPS is classified as a composite supernova remnant \citep{hessco2018} (SNR G338.3-0.0). 
We also note the presence in the region of the pulsar PSR J1640-4631, as well as the presence of SNR G338.5+0.1 \citep{green2025}, both associated with 4FGL sources.
Note that the source HESS~1640-465 studied in \cite{davies2012} is now classified as a SNR and is located in projection at 30 pc from the cluster. 
However, we do find that the unidentified source HESS~J1641-463 is even closer ($\Delta \vartheta/\vartheta_s$=0.76) and therefore potentially associated to this cluster.

\textit{SGR~1806-20:} We find 1 unidentified 4FGL source (J1808.2-2028e, $\Delta \vartheta/\vartheta_s$=0.64) and 1 unidentified HGPS source (HESS J1808-204, $\Delta \vartheta/\vartheta_s$=0.06) within the search region. 
As explained in \cite{Yeung2016}, this region hosts several objects that could produce gamma-rays (the eponymous magnetar, SNR G9.7-0.0, as well as several molecular clouds) but nevertheless the cluster could provide a significant fraction of the observed gamma-rays.

\section{Gamma-ray sources associated with clusters and WR stars}

Tables~\ref{tab:associationsALL} and \ref{tab:associationsALLWR} list sources from 4FGL, HGPS, and First LHAASO Catalogue which are located within a distance $\vartheta_s$ (see main text for the definition of this quantity) from the clusters and WR stars in Fig.~\ref{fig:maps} and \ref{fig:mapsWRs}, respectively.

\clearpage

\begin{table*}
\begin{center}
\caption{\centering List of sources associated with stellar systems}
    \begin{tabular}{| c | c | c | c | c | c | c | c |}
    \hline
    Cluter/{Association~\tablefootmark{~a}} & $\vartheta_s [^{\circ}]$ & 4FGL Name & Type & $\Delta \vartheta / \vartheta_s$ & HGPS/LHAASO Name & Type & $\Delta \vartheta / \vartheta_s$ \\
    \hline
    {Cyg OB1~\tablefootmark{~a}} & 1.15 & 4FGL J2016.2+3712 & SNR & 0.31 & 1LHAASO J2018+3643 & PWN & 0.29 \\
    & & 4FGL J2015.5+3710 & FSRQ & 0.40 & 1LHAASO J2020+3649 & UNID & 0.59 \\
    & & 4FGL J2017.9+3625 & PSR & 0.50 & 1LHAASO J2020+3638 & UNID & 0.60 \\
    & & 4FGL J2021.1+3651 & PSR & 0.61 &  &  &  \\
    & & 4FGL J2022.6+3716c &  & 0.89 &  &  &  \\
    & & 4FGL J2013.5+3613 & SPP & 0.99 &  &  &  \\
    \hline
    {Dragonfish~\tablefootmark{~a}} & 0.57 & 4FGL J1213.3-6240e & SNR & 0.48 &  &  &  \\
    & & 4FGL J1210.4-6250 &  & 0.61 &  &  &  \\
    & & 4FGL J1216.4-6317 &  & 0.86 &  &  &  \\
    \hline
    {Trumpler 27} & 0.79 & 4FGL J1737.3-3332 &  & 0.19 &  &  &  \\
    & & 4FGL J1735.9-3342 & SPP & 0.39 &  &  &  \\
    & & 4FGL J1736.9-3357 & BCU & 0.65 &  &  &  \\
    \hline
    {VVV Cl036} & 0.86 & 4FGL J1409.1-6121e &  & 0.10 &  &  &  \\
    & & 4FGL J1411.5-6133 & PSR & 0.49 &  &  &  \\
    & & 4FGL J1405.1-6119 & HMB & 0.54 &  &  &  \\
    & & 4FGL J1415.3-6110c &  & 0.89 &  &  &  \\
    \hline
    {HM 1} & 0.56 & 4FGL J1719.5-3858c & UNK & 0.35 & HESS J1718-385 & PWN & 0.63 \\
    & & 4FGL J1717.8-3906 &  & 0.63 &  &  &  \\
    & & 4FGL J1718.2-3825 & PSR & 0.75 &  &  &  \\
    \hline
    {Dolidze 33} & 0.63 & 4FGL J1840.0-0411 & SPP & 0.54 &  &  &  \\
    & & 4FGL J1840.8-0453e & SPP & 0.72 &  &  &  \\
    & & 4FGL J1839.2-0420 & SPP & 0.72 &  &  &  \\
    & & 4FGL J1842.5-0359c & UNK & 0.93 &  &  &  \\
    \hline
    {Bochum 14} & 0.63 & 4FGL J1801.6-2326 &  & 0.34 & HESS J1801-233 & SNR & 0.49 \\
    & & 4FGL J1801.3-2326e & SNR & 0.40 & HESS J1800-240 & SNR/MC & 0.78 \\
    & & 4FGL J1801.8-2358 &  & 0.57 &  &  &  \\
    & & 4FGL J1800.7-2355 & UNK & 0.69 &  &  &  \\
    & & 4FGL J1800.9-2407 &  & 0.90 &  &  &  \\
    & & 4FGL J1759.7-2354 &  & 0.94 &  &  &  \\
    & & 4FGL J1800.2-2403c & & 0.96 &  &  &  \\
    \hline
    {Markarian 50} & 0.62 & 4FGL J2317.6+6036c &  & 0.56 &  &  &  \\
    & & 4FGL J2315.9+5955c & UNK & 0.87 &  &  &  \\
    \hline
    {VVV Cl073} & 0.43 & 4FGL J1629.3-4822c &  & 0.55&  &  &  \\
    & & 4FGL J1631.7-4826c &  & 0.73 &  &  &  \\
    & & 4FGL J1631.6-4756e & PWN & 0.82 &  &  &  \\
    \hline
    {Mercer 23} & 0.59 & 4FGL J1929.8+1832 & UNK & 0.17 & HESS J1930+188 & Composite & 0.52 \\
    & & 4FGL J1930.5+1853 & PWN & 0.63 & 1LHAASO J1928+1813u &  UNID & 0.94 \\
    & & 4FGL J1932.4+1846 & SPP & 0.99 &  &  &  \\
    \hline
    {VVV Cl041} & 0.45 & 4FGL J1445.9-5907 &  & 0.62 &  &  &  \\
    & & 4FGL J1444.9-5939 &  & 0.73 &  &  &  \\
    & & 4FGL J1449.8-5923c &  & 0.98 &  &  &  \\
    \hline
    \end{tabular}
\tablefoot{The list includes sources whose angular distance $\Delta \vartheta$ from the system is smaller than $\vartheta_s$ (column 2). Source types are also indicated in columns 4 and 7. The "e" at the end of some of the 4FGL names means that these sources are extended, while the "c" at the end means that they are affected by confusion with the background. 
\tablefoottext{~a}{Stellar associations.}}
\label{tab:associationsALL}
\end{center}
\end{table*}

\clearpage

\begin{table*}[!t]
\begin{center}
\caption{\centering List of sources associated with WR stars from Fig.~\ref{fig:mapsWRs}.}
    \begin{tabular}{| c | c | c | c | c | c | c | c |}
    \hline
    WR & $\vartheta_s [^{\circ}]$ & 4FGL Name & Type & $\Delta \vartheta / \vartheta_s$ & HGPS/LHAASO Name & Type & $\Delta \vartheta / \vartheta_s$ \\
    \hline
    {110} & 1.08 & 4FGL J1808.8-1949c & GLC & 0.44 & HGPSC 056 (HESS J1809-193) & UNID & 0.16 \\
    & & 4FGL J1810.3-1925e & SPP & 0.53 & HGPSC 058 (HESS J1809-193) & UNID & 0.35 \\
    & & 4FGL J1811.5-1925 & PSR & 0.78 & 1LHAASO J1809-1918 & UNID & 0.35 \\
    & & 4FGL J1811.5-1844 & SPP & 0.99 &  HGPSC 059 (HESS J1809-193) & UNID & 0.87 \\
    & & 4FGL J1808.2-2028e & UNID & 1.0 &  HGPSC 057 (HESS J1808-204) & UNID & 0.94 \\
    \hline
    {114} & 0.82 & 4FGL J1823.3-1340 &  & 0.07 & 1LHAASO J1825-1418 & PWN & 0.61 \\
    & & 4FGL J1824.4-1350e & PWN & 0.38 &HGPSC 066 (HESS J1825-137) & PWN & 0.69 \\
    & & 4FGL J1824.1-1304 & SPP & 0.84 & 1LHAASO 1825-1337 & PWN & 0.76 \\
    & &  &  & & HGPSC 067 (HESS J1825-137) & PWN & 0.80 \\
    & &  &  &  & HGPSC 065 (HESS J1825-137) & PWN & 0.81 \\
    \hline
    {111} & 1.04 & 4FGL J1806.2-2126 & spp & 0.53 & HGPSC 057 (HESS J1808-204) & UNID & 0.82 \\
    & & 4FGL J1805.6-2136e & SNR & 0.73 & HGPSC 054 (HESS J1804-216) & UNID & 0.97 \\
    & & 4FGL J1808.2-2028e & UNK & 0.75 &  &  &  \\
    & & 4FGL J1804.7-2144e & SPP & 0.95 &  &  &  \\
    \hline
    {14} & 0.77 & 4FGL J0856.0-4724c &  & 0.33 &  &  &  \\
    & & 4FGL J0851.2-4737 &  & 0.82 &  &  &  \\
    & & 4FGL J0859.2-4729 &  & 0.94 &  &  &  \\
    \hline
    \end{tabular}
\tablefoot{The list includes sources whose angular distance $\Delta \vartheta$ from the system is smaller than $\vartheta_s$ (column 2). Source types are also indicated in columns 4 and 7. The "e" at the end of some of the 4FGL names means that these sources are extended, while the "c" at the end means that they are affected by confusion with the background.}
\label{tab:associationsALLWR}
\end{center}
\end{table*}

\clearpage

\vspace{-10cm}

\section{Chance-coincidence evaluation}

\label{appendix:randomness}

As the Galactic disk is crowded with sources, the possibility that the spatial associations found in Sec.~\ref{sec:rankclusters} and \ref{sec:rankWR} are random coincidences has to be considered.
Here, we limit the analysis to the associations with Fermi-LAT unidentified sources from the 4FGL, as they constitute the large majority of the associations we found.
Also, the 4FGL is the only catalogue of gamma-ray sources among those considered in this work that covers the entire sky, and it contains thousands of sources.
Therefore, it provides a very accurate description of the source distribution in the Galaxy (the HGPS covers the Galactic disk but contains only 78 sources, while the location of LHAASO in the Northern hemisphere prevents it to probe the entire Galaxy).

\begin{figure*}
   \centering
   \includegraphics[width=1\textwidth]{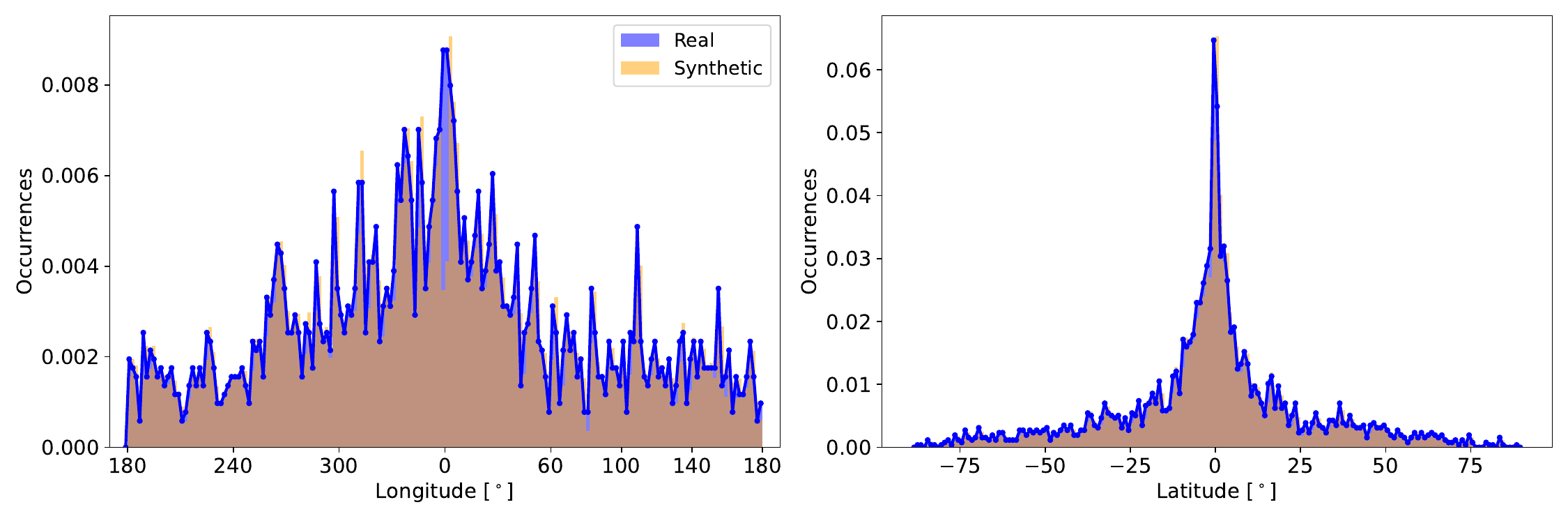}
   \caption{Galactic longitude (left) and latitude (right) distribution of 4FGL unidentified sources (cyan), together with the average distribution obtained from our simulations (orange).}
              \label{fig:distrib_fermi}
    \end{figure*}

We start by sampling the distribution of 4FGL unidentified sources as a function of their longitude and latitude. 
The sample is large enough ($\sim$2500 sources) to show that an overdensity of sources is present towards the Galactic center and also to reveal peaks in the source density corresponding to the Galactic spiral arms \footnote{Note that we added to the sample of 4FGL unidentified sources also the three sources identified as SFR (Star Forming Regions). We did that because a couple of our 44 clusters (Cyg OB2 and Westerlund~2) are associated with one of such sources in 4FGL. This of course has no impact on our simulation procedure, but allows us to reproduce our real matches.}. 
At the same time, the number of sources quickly decreases at large latitudes.
The distributions in Galactic longitude and latitude of the sampled catalog are shown in Fig.~\ref{fig:distrib_fermi}. This constitutes the starting sample from which we develop the Monte Carlo extractions used to evaluate the level of chance coincidence. We realized $10^4$ Monte-Carlo extractions, to generate $10^4$ synthetic catalogs of gamma-ray sources. We overplot one of these $10^4$ realizations on top of the original distributions to demonstrate how these are preserved. 

\begin{figure*}
   \centering
   \includegraphics[width=0.48\textwidth]{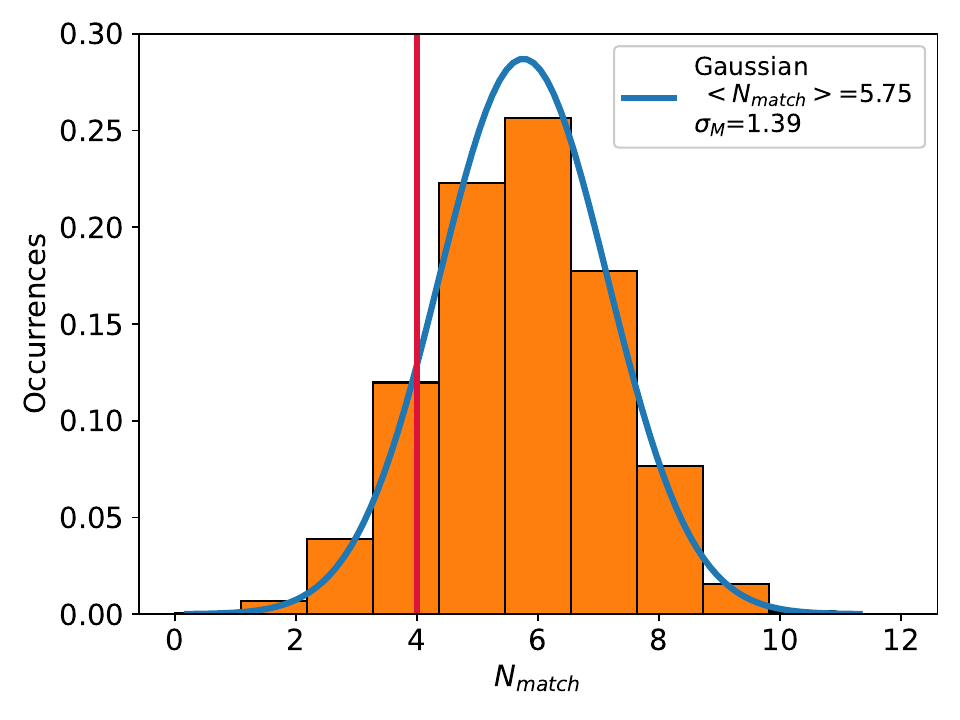}
   \includegraphics[width=0.48\textwidth]{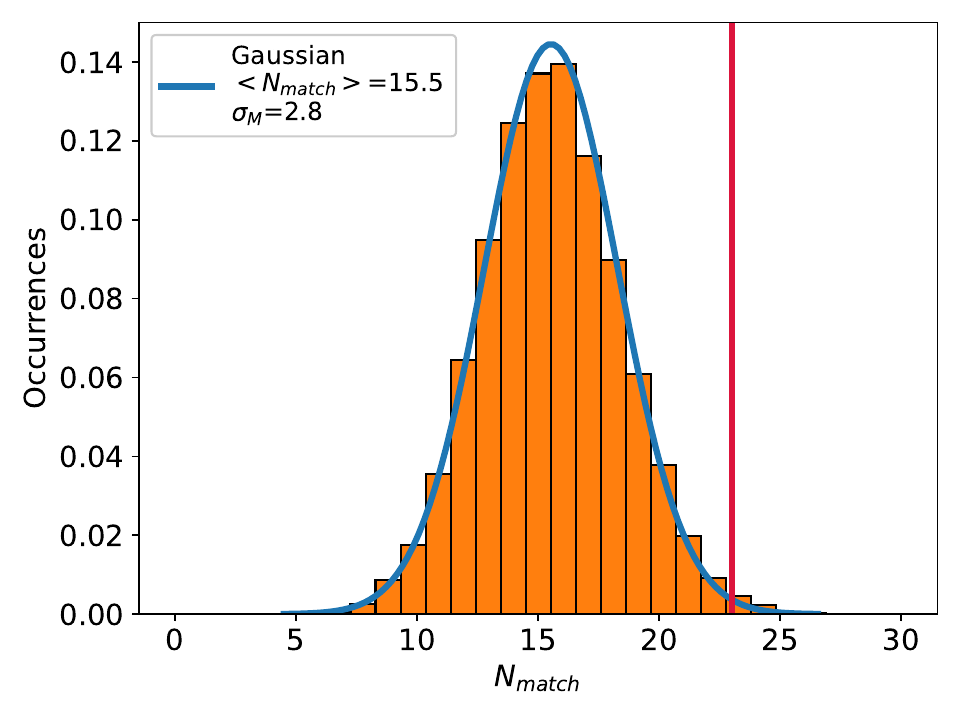}
   \caption{Distribution of the number of matches with the synthetic catalogs and WR stars (left) or star clusters (right). The red vertical line in each plot indicates the number of real matches.The blue line indicates a gaussian distribution with the mean and sigma indicated in the figure legend. }
              \label{fig:distrib_WR_clusters}
    \end{figure*}

We proceed then by counting how many matches we find with the $10^4$ simulated catalogs to evaluate the probability of chance coincidences. 
Similarly to what was done in the paper, the match is defined considering a region of radius $\vartheta_s$ (defined in Sec.~\ref{3.2}) around the 44 star clusters or the 10 WR stars from our samples. 
The results are summarized in Table \ref{tab:sigma_ass}, where we report the number of matching sources found with the real catalog, $N_{match}$, the average number of matches found over the $10^4$ simulated realizations $<N_{match}>$, the standard deviation, $\sigma_{match}$ of the distribution of simulated matches, and the significance of the association evaluated as $\Sigma=(N_{match}-<N_{match}>)/\sigma_{match}$. The quantity $\sigma_{match}$ is the dispersion of the distribution of the number of matches computed as $\sum_i(N_i-<N_{match}>)^2/N$ where $N_i$ are the number of matches found with each one of the N=$10^4$ synthetic catalogs. The resulting distribution is approximately Gaussian,  as one can see in the Fig.~\ref{fig:distrib_WR_clusters}, therefore the above mentioned $\sigma$ can be considered as the analog of a Gaussian-sigma, in that the probability of finding a value with a deviation of 2 sigma from the mean is about 5\% and finding a deviation of more than 3 sigmas is about 0.3\% . For this reason $\Sigma$ gives us an indication on the deviation of our measured value from the mean random value.

Results are reported in Table~\ref{tab:sigma_ass}.
For star clusters, the average number of simulated matches is 15.5, which is smaller than the real matches (which are 23 and include all the 11 new matches found in this paper and shown in Fig.~\ref{fig:maps}, plus the 12 matches with GeV sources previously known, see Table~\ref{tab:rankclusters} and Appendix~\ref{sec:appendix}). 
We found that the significance for an association between star clusters hosting WR stars and 4FGL unidentified sources is $\Sigma \simeq 2.7$. This can be translated in a probability that the associations are random which is smaller than 1\%, providing some evidence against a purely random association.

\begin{figure}
    \centering
    \includegraphics[width=1\linewidth]{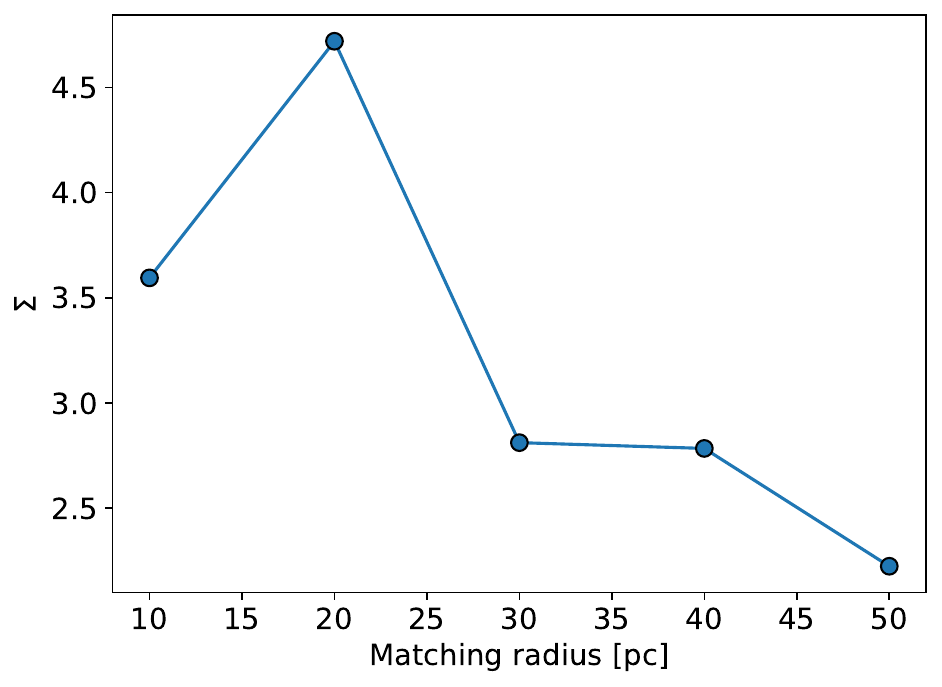}
    \caption{Significance $\Sigma$ of the correlation between star clusters hosting WR stars and unidentified 4FGL sources, as defined in the main text, derived for different matching radii.}
    \label{fig:sigma_radii}
\end{figure}

As mentioned in the main text, we tested the stability of our results by trying different search radii.
We varied the size of the region of interest $R_s$ in the range 10 to 50~pc with a step of 10~pc.
The significance of the correlation $\Sigma$ is shown in Fig.~\ref{fig:sigma_radii}, and is characterized by a peak around 20~pc. Using this radius we get $\Sigma \simeq 4.7$, corresponding to a probability that these matches are random smaller than 10$^{-4}$, i.e., with such a number of simulations, we never find a value larger than 19, the real number of matches obtained with a 20-pc matching radius.

From our simulations, we can also estimate the number of spatial associations which are not random.
Taking the results of the simulations performed with a fixed search region of 20~pc, we find that the number of associations which are not random should be $10.8 \pm 2.3$.
It is interesting to note that this number is significantly larger than the number of clusters in our sample (list in Table~\ref{tab:rankclusters}) which are associated with a 4FGL source classified as Star Forming Regions (two objects: Cygnus and Westerlund~2), and even larger than the number of clusters for which we estimated that an association with a GeV gamma-ray source is ``Likely'' (5 objects: see Table~\ref{tab:rankclusters}).
These figures are consistent with our claim of a hint for a correlation.

On the other hand, this cannot be said for WRs, since the resulting $\Sigma$ is negative, suggesting that most cases have a large probability to be random coincidences.
This does not mean, of course, that none of the spatial associations is real, but that these objects, as a population, cannot be considered gamma-ray emitters detectable by Fermi-LAT.
As stressed in the main body of the manuscript, we have considered here only a quite small number of WR stars that satisfy a necessary condition to be gamma-ray sources, i.e., they are characterised by a large value of $L_w/d^2$.

This necessary condition is satisfied by very few objects, but these are the only ones from which one can hope to detect gamma-rays, and it is therefore important to study them.
A sufficient condition to have gamma-ray emission would be to have a WR star of large $L_w/d^2$ located in an environment which is favorable to the production of gamma rays. 
As environmental conditions are very uncertain, it seems appropriate to better investigate these very few objects, both searching for gamma-ray emission from their immediate surroundings (as we did in this paper), but also trying to characterise better their ambient conditions. 
If ambient conditions are well constrained and found to be favorable, even a non detection would be interesting as it will constrain the CR acceleration efficiency at WTSs.
It seems, then, that these very few objects deserve a case-by-case study.

A case-by-case study can be useful also for star clusters, as it might help in prioritising future observations or analyses of such objects. 
With this respect, we note that in a few cases, the number of unidentified sources within the search region, $N_{unid}$, is larger than one. 
This is particularly interesting as the diffuse emission associated with some star clusters (tentatively or likely, see Table~\ref{tab:rankclusters}) was in fact seen in the 4FGL as a number of discrete sources scattered in the region.
A criterion is needed to decide whether the concentration of sources ($N_{unid} > 1$) seen around some clusters are significant or just compatible with fluctuations.

Before illustrating our case-by-case study, an important remark is in order.
The gamma-ray sources and the stellar objects considered here lie in the Galactic disk, where their spatial distribution is far from homogeneous on small spatial scales, and large fluctuations in source density over angular scales of order of $\approx 1^{\circ}$ or so are common.
Therefore, considering only quantities as  the average number of unidentified 
gamma-ray sources expected within the search radius in comparison with the observed $N_{unid}$ can lead to biased conclusions. One should also assess the local source density around each object, 
in order to understand whether a  $N_{unid}$ larger (smaller) than average represents a true local excess (deficit) or results from a region more (less) crowded than average.

To do so, let us consider two regions around each WR or SC, a inner (dubbed "in") coincident with the rearch region of radius $\vartheta_s$, and a ring (dubbed "ring") with outer radius twice that of the inner region. 
The ratio of area of the inner and ring regions is then $A_{ring}/A_{in}= 3 $.
We call the number of sources within the inner and ring region $N_{in} \equiv N_{unid}$ and $N_{ring}$, respectively.
An overdensity of sources with respect to the local density is therefore defined by the condition $N_{in}/N_{ring} > 1/3$ or, equivalently, by a local source excess $\delta = N_{in} - N_{ring}/3 > 0$. We tested for potential biases in this method and found that neither the the choice of the matching radius nor the gradient in the spatial distribution of gamma-ray sources affect significantly $\delta$, since the latitude distribution of Fermi/LAT sources is much broader than that of the star clusters or WRs.

We present in Table.\ref{tab:statistics} the values of $\delta$ for the 23 studied clusters and we note that 17 over 23 respect this condition. In particular, the ones characterized by the largest values of $N_{in}/N_{out}$ are  Bochum 14 ($\delta = 4.67$); Mercer 20 and VVV Cl041 ($\delta = 2.67$). Concerning the WR stars considered here, the values of $\delta$ are also reported in Table.\ref{tab:statistics}, and we get that 3 over 4 satisfy this condition: WR~14, 110 and 111. 

To further investigate those objects, we computed for each of them the fraction of simulations in which we find a number of unidentified sources larger or equal than $N_{unid}$ within the search region. A large fraction would indicate that it is likely that the associations are random. These fractions are also reported in Table.\ref{tab:statistics} and they are smaller than 1~\% for the star clusters Bochum~14, VVV~Cl041, Mercer~20, Danks~1, and W43, while it is about 2\% for WR~14. We note in addition (see Fig.~\ref{fig:maps} and \ref{fig:mapsWRs}) that the search regions surrounding VVV~CL041 and WR~14 are particularly clean: they both contain 3 unidentified 4FGL sources and no identified ones. From these fractions, random-chance coincidence probabilities can then be computed, taking into account trials (we searched for gamma-ray sources around 44 clusters and 10 WR stars). For small values of $F(\geq N_{\rm unid, ~4FGL})$ the probability can be simply obtained by multiplying this number times the number of trials. The smallest value of $F(\geq N_{\rm unid, ~4FGL})$ have been obtained for Mercer~20 and Bochum~14, and would correspond to a random-chance coincidence probability at the few percent level.

\begin{table}[]
    \centering
    \caption{\centering Number of spatial associations $N_{match}$ with 4FGL unidentified sources found within an angular distance $\vartheta_s$ (see Sec.\ref{3.2}) from clusters hosting WR stars (Table \ref{tab:rankclusters}) and bright WRs (Table \ref{tab:WRs}).}
    \begin{tabular}{l|c | >{}c  | >{}c |  > {}c |}
       Sample  & $N_{match}$ &  $<N_{match}>$ & $\sigma_{match}$ & $\Sigma$  \\
       \hline
        Star Clusters & 23 & 15.5 & 2.80  & 2.7 \\
        Star Clusters [20~pc]  & 19 & 8.17 & 2.29  & 4.7 \\
        Bright WRs & 4 & 5.75 & 1.39 & -1.3 \\
    \end{tabular}
    \tablefoot{The second row shows the results obtained for a fixed search region of size 20~pc. The third, fourth, and fifth columns shown the number of mean simulated matches, their dispersion, and their significance.}
    \label{tab:sigma_ass}
\end{table}

\begin{table*}
\begin{center}
\caption{\centering Excesses of local sources and simulated fractions of random spatial coincidence for WR stars and star clusters/associations.}
    \begin{tabular}{| c | c | c | c | >{}c |}
    \hline
    Cluter/{Association~\tablefootmark{~a}}/WR & $N_{\rm {unid, ~4FGL}} $ & $N_{\rm {unid, ~4FGL}}^{ring} $ & $\delta$ & $F(\geq N_{\rm unid, ~4FGL})$ \\
    \hline
    Westerlund 1 & 2 & 2 & 1.333 & 0.2203 \\ \hline
    Dragonfish~\tablefootmark{~a} & 1 & 1 & 0.667 & 0.3796 \\ \hline
    Danks 1 & 3 & 3 & 2.000 & 0.005 \\ \hline
    Car OB1~\tablefootmark{~a} & 1 & 2 & 0.333 & 0.2833 \\ \hline
    NGC 3603 & 1 & 1 & 0.667 & 0.0497 \\ \hline
    SGR 1806 20 & 1 & 0 & 1.000 & 0.0706 \\ \hline
    Cyg OB1~\tablefootmark{~a} & 1 & 4 & -0.333 & 0.7208 \\ \hline
    Cyg OB2~\tablefootmark{~a} & 1 & 2 & 0.333 & 0.5831 \\ \hline
    Mercer 20 & 4 & 4 & 2.667 & 0.0006 \\ \hline
    VVV Cl073 & 2 & 1 & 1.667 & 0.1032 \\ \hline
    Westerlund 2 & 2 & 0 & 2.000 & 0.0124 \\ \hline
    HM 1 & 2 & 2 & 1.333 & 0.0776 \\ \hline
    Trumpler 27 & 1 & 7 & -1.333 & 0.7233 \\ \hline
    VVV Cl041 & 3 & 1 & 2.667 & 0.0059 \\ \hline
    VVV Cl036 & 2 & 17 & -3.667 & 0.6748 \\ \hline
    Danks 2 & 2 & 3 & 1.000 & 0.0157 \\ \hline
    Gamma Vel~\tablefootmark{~a} & 9 & 56 & -9.667 & 0.8647 \\ \hline
    W43 & 2 & 2 & 1.333 & 0.0061 \\ \hline
    Dolidze 33 & 1 & 3 & 0.000 & 0.7372 \\ \hline
    Markarian 50 & 2 & 2 & 1.333 & 0.1197 \\ \hline
    Mercer 23 & 1 & 3 & 0.000 & 0.3697 \\ \hline
    Berkeley 87 & 2 & 2 & 1.333 & 0.2283 \\ \hline
    Bochum 14 & 6 & 4 & 4.667 & 0.0008 \\ \hline\hline
    WR 110 & 1 & 2 & 0.333 & 0.9236 \\ \hline
    WR 114 & 1 & 3 & 0.000 & 0.7655 \\ \hline
    WR 111 & 1 & 0 & 1.000 & 0.8227 \\ \hline
    WR 14 & 3 & 3 & 2.000 & 0.0222 \\ \hline
    \end{tabular}
\tablefoot{In particular, for each object, we report the number of unidentified sources in the inner region, in the ring region, as well as the $\delta$ parameter. We also give the fraction of simulations where we have at least the same number of coincident unidentified sources compared to the real sample, calculated as explained in the text. Random chance coincidence probabilities taking into account a trial factor can then be computed as described in the text.
\tablefoottext{~a}{Stellar associations}
}
\label{tab:statistics}
\end{center}
\end{table*}

\end{appendix}

\end{document}